\RequirePackage{lineno}
\documentclass[twocolumn,aps,tightenlinefs,superscriptaddress,preprintnumbers]{revtex4}
\usepackage{graphicx,color,dcolumn,booktabs,bm}
\usepackage{longtable,lscape}
\usepackage{amsmath}
\usepackage{indentfirst}
\usepackage{epsfig}
\usepackage{feynmf}
\usepackage{epstopdf}
\usepackage{slashed}
\usepackage{cases}
\usepackage{color}
\usepackage{multirow}
\usepackage{ulem}
\usepackage{graphicx,color,dcolumn,booktabs,bm}
\usepackage{verbatim}
\usepackage[colorlinks,linkcolor=black,anchorcolor=green,citecolor=blue]{hyperref}
\usepackage{mathrsfs}
\usepackage{rotating}
\usepackage{threeparttable}
\usepackage{dcolumn}
\usepackage{orcidlink} 

\newcommand{\ppp}{{\cal P}}
\newcommand{\eff}{\varepsilon}
\newcommand{\BR}{{\cal B}}

\newcommand{\pip}{\pi^+}
\newcommand{\pim}{\pi^-}

\newcommand{\EE}{e^+e^-}

\newcommand{\F}{{\cal F}}

\newcommand{\beq}{\begin{equation}}
\newcommand{\eeq}{\end{equation}}
\newcommand{\bitm}{\begin{itemize}}
\newcommand{\eitm}{\end{itemize}}

\newcommand{\fours}{\Upsilon(4S)}
\newcommand{\fives}{\Upsilon(5S)}

\begin{document}
\hyphenpenalty=10000


\noaffiliation
  \author{Y.~Li\,\orcidlink{0000-0002-4413-6247}} 
  \author{C.~P.~Shen\,\orcidlink{0000-0002-9012-4618}} 
  \author{I.~Adachi\,\orcidlink{0000-0003-2287-0173}} 
  \author{H.~Aihara\,\orcidlink{0000-0002-1907-5964}} 
  \author{D.~M.~Asner\,\orcidlink{0000-0002-1586-5790}} 
  \author{V.~Aulchenko\,\orcidlink{0000-0002-5394-4406}} 
  \author{T.~Aushev\,\orcidlink{0000-0002-6347-7055}} 
  \author{V.~Babu\,\orcidlink{0000-0003-0419-6912}} 
  \author{P.~Behera\,\orcidlink{0000-0002-1527-2266}} 
  \author{K.~Belous\,\orcidlink{0000-0003-0014-2589}} 
% ...
\collaboration{The Belle Collaboration}

\title{\quad\\[0.1cm]\boldmath First search for the weak radiative decays
$\Lambda_c^+ \to \Sigma^+ \gamma$ and $\Xi_c^0 \to \Xi^0 \gamma$}

\begin{abstract}
We present the first search for the weak radiative decays $\Lambda_c^+ \to \Sigma^+ \gamma$
and $\Xi_c^0 \to \Xi^0 \gamma$ using a data sample of 980~fb$^{-1}$ collected by the
Belle detector operating at the KEKB asymmetric-energy $\EE$ collider. There are no
evident $\Lambda_c^+ \to \Sigma^+ \gamma$ or $\Xi_c^0 \to \Xi^0 \gamma$ signals. Taking the
decays $\Lambda_c^+ \to p K^- \pip$ and $\Xi_c^0 \to \Xi^- \pip$  as normalization channels,
the upper limits at 90\% credibility level on the ratios of branching fractions
$\BR(\Lambda_c^+ \to \Sigma^+ \gamma)/\BR(\Lambda_c^+ \to p K^{-} \pip) < 4.0 \times 10^{-3}$ and
$\BR(\Xi_c^0 \to \Xi^0 \gamma)/\BR(\Xi_c^0 \to \Xi^- \pi^+) < 1.2 \times 10^{-2}$ are determined.
We obtain the upper limits at 90\% credibility level on the absolute branching fractions
$\BR(\Lambda_c^+ \to \Sigma^+ \gamma) < 2.6 \times 10^{-4}$ and $\BR(\Xi_c^0 \to \Xi^0 \gamma) < 1.8 \times 10^{-4}$.
\end{abstract}

\maketitle
\section{\boldmath Introduction}
Charm physics has always been a popular topic due to the fact that the
charm system provides a distinctive laboratory to investigate the interplay
of strong and weak interactions. Weak radiative decays of charmed
hadrons proceed via $W$-exchange, and are dominated by long-distance
nonperturbative processes; short-distance contributions from electromagnetic
penguin diagrams are highly suppressed~\cite{Burdman:1995te,Greub:1996wn}.
The long-distance contributions to the Cabibbo-favored (CF) weak radiative decays of charmed hadrons
are predicted to have branching fractions at the level of $10^{-4}$~\cite{Burdman:1995te,Greub:1996wn,Fajfer:1997bh,Uppal,Kamal,Cheng,Bajc:1994ui,Fajfer:2015zea}.
Measurements of the branching fractions of weak radiative decays of charmed hadrons can be used to test
long-distance dynamics calculations based on different theoretical models.

In the charmed meson sector, several weak radiative decays have been reported~\cite{Belle:2003vsx,BaBar:2008kjd,Belle:2016mtj}.
The Cabibbo-suppressed (CS) weak radiative decay $D^0 \to \phi \gamma$ was first observed by the Belle experiment~\cite{Belle:2003vsx}.
The {\textsl{BABAR}} experiment found the CF weak radiative decay $D^0 \to \bar{K}^{*}(892)^{0}\gamma$~\cite{BaBar:2008kjd}.
In 2017, the Belle experiment presented the first observation of the CS weak radiative decay $D^0 \to \rho^0 \gamma$ with a measured branching
fraction $\BR(D^0 \to \rho^0 \gamma) = (1.77\pm0.30\pm0.07)\times10^{-5}$ and the improved
measurements of branching fractions $\BR(D^0 \to \phi \gamma) = (2.76\pm0.19\pm 0.10)\times10^{-5}$
and $\BR(D^0 \to \bar{K}^{*}(892)^{0}\gamma) = (4.66\pm0.21\pm0.21)\times10^{-4}$~\cite{Belle:2016mtj},
where the first and second uncertainties are statistical and systematic, respectively. However, the weak
radiative decays of charmed baryons have not yet been measured.

\begin{figure}[htbp]
	\begin{center}
		\includegraphics[width=4.55cm]{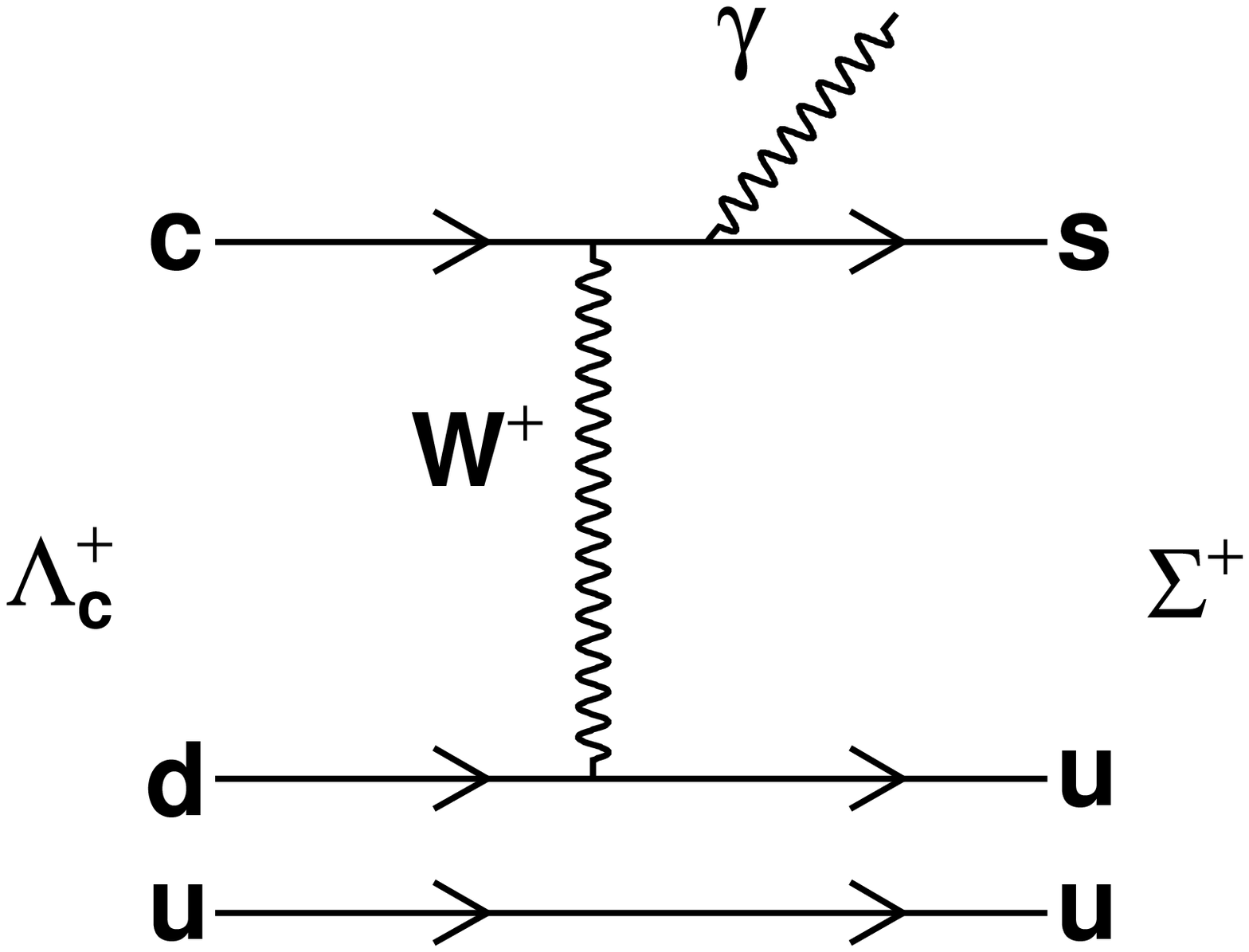}\hspace{-0.55cm}
		\includegraphics[width=4.55cm]{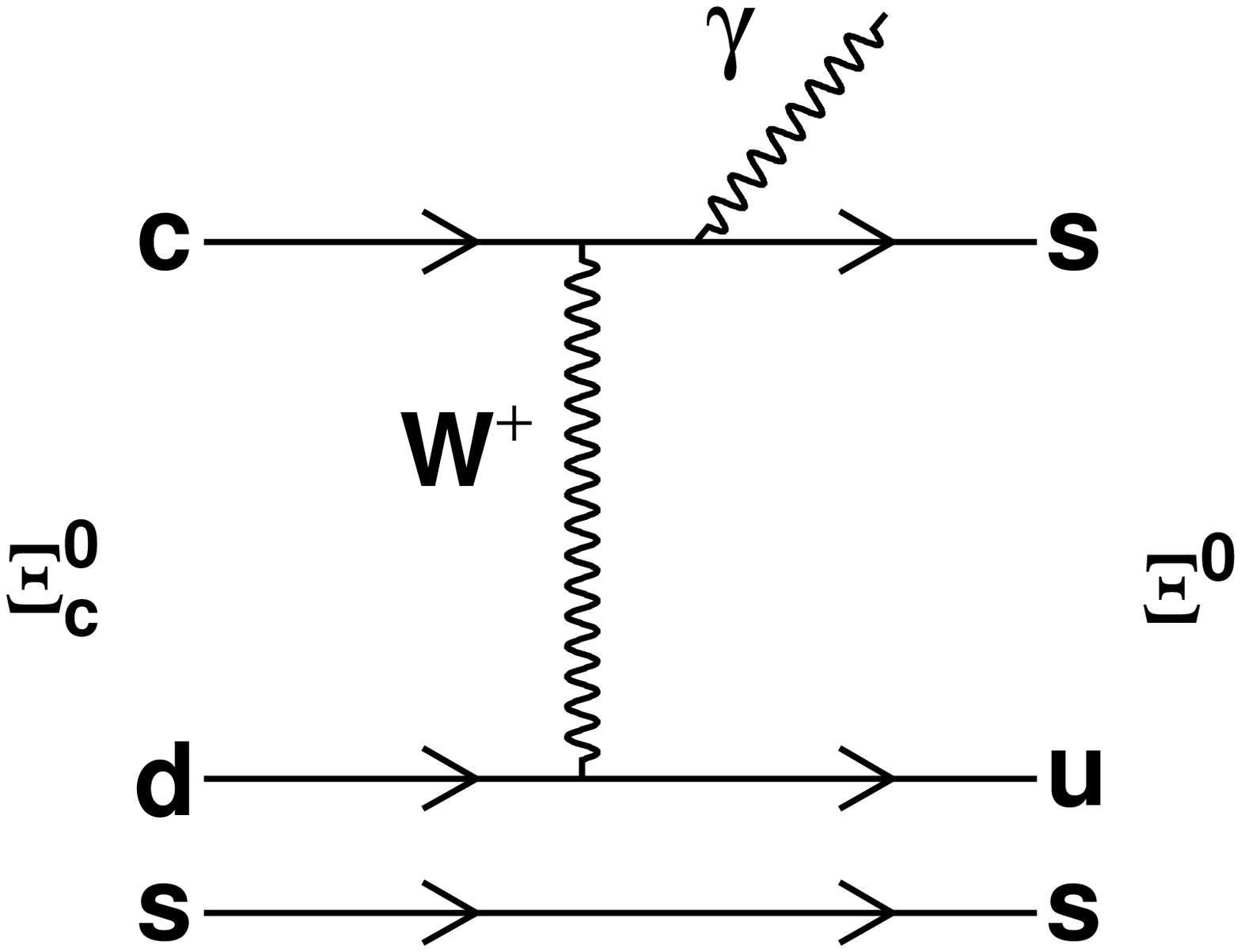}
		\put(-235,80){\bf (a)} \put(-120,80){\bf (b)}
		\caption{Examples of $W$-exchange diagrams accompanied by photon emission from the external $s$ quark for
		(a)	$\Lambda_c^+ \to \Sigma^+ \gamma$ and (b) $\Xi_c^0 \to \Xi^0 \gamma$ decays.}\label{Fig1}
	\end{center}
\end{figure}

The LHCb experiment observed the first weak radiative decay of a bottom baryon $\Lambda_b^0 \to \Lambda \gamma$
in 2019, and measured the branching fraction $\BR(\Lambda_b^0 \to \Lambda \gamma) =
(7.1 \pm 1.5 \pm 0.6 \pm 0.7) \times 10^{-6}$, where the quoted uncertainties are statistical,
systematic, and from the external inputs, respectively~\cite{LHCb_Lbb}. The decay
$\Lambda_b^0 \to \Lambda \gamma$ proceeds via the $b \to s \gamma$ flavor-changing neutral-current
transition, which is dominated by short-distance processes. Since the penguin process
$c \to u \gamma$ is highly suppressed, it plays very little role in the weak radiative decays of charmed baryons.
More dominant contributions to the weak radiative decays of charmed baryons could arise from
$W$-exchange bremsstrahlung processes such as $c d \to u s \gamma$. The $c d \to u s \gamma$ process induces
two CF weak radiative decays of anti-triplet charmed baryons: $\Lambda_c^+ \to \Sigma^+ \gamma$
and $\Xi_c^0 \to \Xi^0 \gamma$~\cite{Cheng:2021qpd}. Figure~\ref{Fig1} shows the $W$-exchange
diagrams accompanied by a photon emission from the external $s$ quark for $\Lambda_c^+ \to \Sigma^+ \gamma$ and
$\Xi_c^0 \to \Xi^0 \gamma$ decays as examples. The same $W$-exchange diagrams, but with a photon radiated from other external
quarks, can also contribute to the weak radiative decays $\Lambda_c^+ \to \Sigma^+ \gamma$ and $\Xi_c^0 \to \Xi^0 \gamma$~\cite{Kamal}.
The branching fractions of the decays $\Lambda_c^+ \to \Sigma^+ \gamma$ and $\Xi_c^0 \to \Xi^0 \gamma$
were predicted by the different theoretical methods, including a modified nonrelativistic quark model~\cite{Kamal},
the constituent quark model~\cite{Uppal}, and the effective Lagrangian approach~\cite{Cheng}.
Theoretical branching fraction estimates cover ranges of $(4.5 - 29.1) \times 10^{-5}$ and $(3.0 - 19.5) \times 10^{-5}$
for $\Lambda_c^+ \to \Sigma^+ \gamma$ and $\Xi_c^0 \to \Xi^0 \gamma$ decays~\cite{Kamal,Uppal,Cheng}, respectively, as listed in
Table~\ref{tab:Table1}. There are two estimates in Ref.~\cite{Uppal}, and the case (II) naively considered
the flavor dependence of charmed baryon wave-function squared at the origin $|\psi(0)|^{2}$.
Very recently, the authors of Ref.~\cite{Adolph:2022ujd} propose a new method of
self-analyzing final states to test the standard model and search for possible new physics (NP) in
the radiative decays of charmed baryons. Measuring the branching fractions of weak radiative decays
$\Lambda_c^+ \to \Sigma^+ \gamma$ and $\Xi_c^0 \to \Xi^0 \gamma$ can not only exclude the parameter space of
NP existence but also yield experimental inputs for the  theoretical understanding of long-distance interactions
in the weak radiative decays of charmed hadrons.

\begin{table}[htbp]
	\caption{\label{tab:Table1} Theoretical estimates of branching fractions in units of $10^{-5}$ for
		the CF weak radiative decays $\Lambda_c^+ \to \Sigma^+ \gamma$ and $\Xi_c^0 \to \Xi^0 \gamma$. There
		are two predictions in Ref.~\cite{Uppal} depending on the evaluation of the charmed baryon wave-function
		squared at the origin $|\psi(0)|^{2}$. The branching fractions have been rescaled based
		on the current lifetimes of $\Lambda_c^+$ and $\Xi_c^0$ by the author of Ref.~\cite{Cheng:2021qpd}.}
	\begin{tabular}{lcccc}
		\hline\hline
		\multicolumn{1}{c}{Modes} &  \multicolumn{1}{c}{Kamal~\cite{Kamal}} & \multicolumn{2}{c}{Uppal~\cite{Uppal}} &
		\multicolumn{1}{c}{Cheng~\cite{Cheng}} \\
		& & (I) & (II) & \\
		\hline
		$\Lambda_c^+ \to \Sigma^+ \gamma$ &  6.0  & 4.5 & 29.1   & 4.9  \\
		$\Xi_c^0 \to \Xi^0 \gamma$        &  ...  & 3.0 & 19.5   & 4.8  \\
		\hline\hline
	\end{tabular}
\end{table}

In this paper, we perform the first search for the weak radiative decays $\Lambda_c^+ \to \Sigma^+ \gamma$
and $\Xi_c^0 \to \Xi^0 \gamma$ using the entire data sample of 980~fb$^{-1}$ collected
by the Belle detector. The decays $\Lambda_c^+ \to p K^- \pip$ and $\Xi_c^0 \to \Xi^- \pip$ are taken
as normalization channels. Charge-conjugate modes are also implied unless otherwise stated throughout
this paper.

\section{\boldmath The data sample and the belle detector}
This analysis is based on data collected at or near the $\Upsilon(nS)$ $(n = 1,~2,~3,~4,~5)$ resonances
by the Belle detector~\cite{detector1, detector2} at the KEKB asymmetric-energy $e^+e^-$
collider~\cite{collider1, collider2}. The total data sample corresponds to an integrated
luminosity of 980~fb$^{-1}$~\cite{detector2}. The Belle detector is a large-solid-angle
magnetic spectrometer consisting of a silicon vertex detector, a 50-layer central drift chamber (CDC),
an array of aerogel threshold Cherenkov counters (ACC), a barrel-like arrangement of time-of-flight
scintillation counters (TOF), and an electromagnetic calorimeter comprising CsI(Tl)
crystals (ECL) located inside a superconducting solenoid coil that provides a $1.5~\hbox{T}$
magnetic field. An iron flux return comprising resistive plate chambers located outside the
coil is instrumented to detect $K^{0}_{L}$ mesons and to identify muons. The detector is
described in detail elsewhere~\cite{detector1, detector2}.

Monte Carlo (MC) simulated signal events are generated using {\sc EvtGen}~\cite{evtgen}
to optimize the signal selection criteria and calculate the reconstruction efficiencies.
Events for the  $\EE \to c\bar{c}$ production are generated using {\sc PYTHIA}~\cite{pythia}
with a specific Belle configuration, where one of the two charm quarks hadronizes into a
$\Lambda_c^+$ or $\Xi_{c}^0$ baryon. The decays $\Lambda_c^+ \to \Sigma^+ \gamma$,
$\Xi_c^0 \to \Xi^0 \gamma$, $\Lambda_c^+ \to pK^-\pi^+$, $\Xi_c^0\to \Xi^-\pip$,
$\Lambda_c^+ \to \Sigma^+\pi^0$, $\Lambda_c^+ \to \Sigma^+\eta$, $\Xi_c^0 \to \Xi^0\pi^0$,
and $\Xi_c^0 \to \Xi^0 \eta$ are generated using a phase space model.
The simulated events are processed with a detector simulation based on {\sc GEANT3}~\cite{geant}.
Inclusive MC samples of $\Upsilon(1S,~2S,~3S)$ decays, $\fours \to B^{+}B^{-}/B^{0}\bar{B}^{0}$,
$\fives \to B_{(s)}^{(*)} \bar{B}_{(s)}^{(*)}$, and $\EE \to q\bar{q}$ ($q=u,\,d,\,s,\,c$)
at center-of-mass (C.M.) energies of 9.460, 10.024, 10.355, 10.520, 10.580, and 10.867~GeV
corresponding to two times the integrated luminosity of data are used to check for possible
peaking backgrounds and optimize the signal selection criteria.

\section{\boldmath Common Event selection criteria \label{secII}}
We reconstruct the decays $\Lambda_c^+ \to \Sigma^+ \gamma$, $\Xi_c^0 \to \Xi^0 \gamma$,
$\Lambda_c^+ \to p K^- \pi^+$, and $\Xi_c^0 \to \Xi^- \pi^+$. The $\Sigma^+$, $\Xi^0$, and $\Xi^-$
hyperons are reconstructed via $\Sigma^+ \to p \pi^0$, $\Xi^0 \to \Lambda \pi^0$, and $\Xi^- \to \Lambda \pi^-$
decays with the $\pi^0$ and $\Lambda$ in $\pi^0 \to \gamma\gamma$ and $\Lambda \to p \pi^-$ decays, respectively.
The event selection criteria described below are optimized by maximizing the figure-of-merit
$\epsilon/(3/2 + \sqrt{N_{\rm bkg}})$~\cite{Punzi}, where $\epsilon$ is the signal
reconstruction efficiency of $\Lambda_c^+ \to \Sigma^+ \gamma$ or $\Xi_c^0 \to \Xi^0 \gamma$
decay, and $N_{\rm bkg}$ is the number of estimated background events from the
normalized inclusive MC samples in the $\Lambda_c^+$ or $\Xi_c^0$ signal region
defined as 2.18~GeV/$c^2$ $<$ $M(\Sigma^+ \gamma)$ $<$ 2.34~GeV/$c^2$
or 2.36~GeV/$c^2$ $<$ $M(\Xi^0 \gamma)$ $<$ 2.52~GeV/$c^2$ ($>$ 95\% signal events are
retained according to signal MC simulations), respectively. Hereinafter, $M$ represents the measured invariant mass.

For the particle identification (PID) of a well-reconstructed charged track,
information from different detector subsystems, including specific ionization in the CDC, time
measurement in the TOF, and the response of the ACC, is combined to form a
likelihood ratio, $\mathcal{R}(h|h^{\prime}) = \mathcal{L}(h)/[\mathcal{L}(h) + \mathcal{L}(h^{\prime})]$,
where $\mathcal{L}(h^{(\prime)})$ is the likelihood of the charged track being a hadron $h^{(\prime)}$,
and $h^{(\prime)}$ is $p$, $K$, or $\pi$ as appropriate~\cite{pidcode}.
To identify the proton used in $\Sigma^+$ reconstruction, we require
$\mathcal{R}(p|K) > 0.6$ and $\mathcal{R}(p|\pi) > 0.6$, which has an efficiency of 97\%;
we also require a momentum above 0.9~GeV/$c$ in the laboratory frame.
For the proton used in $\Lambda$ reconstruction, we require $\mathcal{R}(p|K) > 0.2$
and $\mathcal{R}(p|\pi) > 0.2$ with an efficiency of 98\%.

An ECL cluster is taken as a photon candidate if it does not match the extrapolation of any
charged track. The $\pi^0$ candidates used in $\Sigma^+$ ($\Xi^0$) reconstruction are formed
from two photons having energy exceeding 50~MeV (30~MeV) in the barrel ($-0.63 < \cos\theta < 0.85$) or 70~MeV (50~MeV)
in the endcaps ($-0.91 < \cos\theta < -0.63$ or $0.85 < \cos\theta < 0.98$) of the ECL, where $\theta$
is the polar angle relative to the opposite direction of $e^+$ beam. The reconstructed
invariant mass of the $\pi^0$ candidate is required to be within 10.8~MeV/$c^2$ of the $\pi^0$ nominal mass~\cite{PDG}, corresponding to approximately
twice the mass resolution ($\sigma$). To reduce the large combinatorial backgrounds,
the momentum of the $\pi^0$ used in $\Sigma^+$ ($\Xi^0$) reconstruction is required to exceed 300~MeV/$c$ (200~MeV/$c$)
in the laboratory frame. The $\Lambda$ candidates are reconstructed in the decay $\Lambda \to p \pi^-$ and selected if
$|M(p\pim)-m(\Lambda)|<$ 3.5~MeV/$c^2$ ($\sim$2.5$\sigma$). Here and throughout this paper,
$m(i)$ represents the nominal mass of the particle $i$~\cite{PDG}.

The $\Sigma^+ \to p \pi^0$ and $\Xi^0 \to \Lambda \pi^0$ reconstructions are complicated
by the fact that the parent hyperon decays with a $\pi^0$, which has negligible vertex
position information, as one of its daughters. For the $\Sigma^+ \to p \pi^0$
reconstruction, combinations of $\pi^0$ candidates and protons are made using those
protons with a sufficiently large ($>$ 1~mm) distance of closest approach to the
interaction point (IP). Then, taking the IP as the point of origin of the $\Sigma^+$,
the sum of the proton and $\pi^0$ momenta is taken as the momentum vector of the $\Sigma^+$
candidate. The intersection of this trajectory with the reconstructed proton trajectory
is then found and this position is taken as the decay location of the $\Sigma^+$ hyperon.
The $\pi^0$ is then re-made from the two photons, using this location as its point of origin.
Only those combinations with the decay location of the $\Sigma^+$ indicating a positive $\Sigma^+$
path length are retained. The $\Xi^0 \to \Lambda \pi^0$ decays are reconstructed using a similar method,
and only those combinations with the decay location of the $\Xi^0$ indicating a positive $\Xi^0$
path length of greater than 2~cm but less than the distance between the $\Lambda$ decay vertex
and the IP are retained~\cite{Belle:2017szm}.

The following criteria are used to select the radiative photon candidates. The energy
of the photon is required to exceed 0.65~GeV in the barrel or 0.8~GeV
in the endcaps of the ECL. To reduce photon candidates originating from neutral hadrons,
we reject a photon candidate if the ratio of energy deposited in the central $3 \times 3$ square
of cells to that deposited in the enclosing $5 \times 5$ square of cells in its ECL cluster is
less than 0.95. Most background photons originate from  $\pi^0 \to \gamma \gamma$ and $\eta \to \gamma \gamma$
decays. To reduce such backgrounds, probability functions are employed to
distinguish the radiative photon candidates from $\pi^0$ and $\eta$ decays.
We first combine the photon candidate with all other photons and calculate likelihoods for the reconstructed photon pair to be
$\pi^0$-like [$\ppp(\pi^0)$] and $\eta$-like [$\ppp(\eta)$], by using the invariant mass of the photon
pair along with the energy of the candidate photon in the laboratory frame and the angle with respect to the beam direction in the
laboratory frame~\cite{veto}. The background photons originating from $\pi^0 \to \gamma \gamma$
and $\eta \to \gamma \gamma$ decays are suppressed by requiring $\ppp(\pi^0) < 0.3$
and $\ppp(\eta) < 0.3$.

The $\Sigma^+ \gamma$ and $\Xi^0 \gamma$ combinations are made to form $\Lambda_c^+$ and $\Xi_c^0$
candidates, respectively. To reduce the combinatorial backgrounds, especially from $B$-meson decays,
the scaled momentum $x_{p} = p^{*}_{i}$/$p_{\rm max}$ is required to be larger than 0.55.
Here, $p^{*}_{i}$ is the momentum of the $\Lambda_c^+$ or $\Xi_{c}^0$ candidate in the $\EE$ C.M.\ frame,
and $p_{\rm max}=\frac{1}{c}\sqrt{E^2_{\rm beam}-M_{i}^2 c^4}$, where $E_{\rm beam}$ is the
beam energy in the $\EE$ C.M.\ frame and $M_{i}$ represents the invariant mass of the $\Lambda_c^+$ or $\Xi_{c}^0$
candidate.

For the normalization channels $\Lambda_c^+ \to p K^- \pip$ and
$\Xi_c^0 \to \Xi^-\pip$, the selection criteria are similar to
those used in Refs.~\cite{Belle:2021vyq, Belle:2021zsy} and are described below.
For the $\Lambda_c^+ \to pK^-\pi^+$ reconstruction, tracks having
$\mathcal{R}(p|K)$ $>$ 0.9 and
$\mathcal{R}(p|\pi)$ $>$ 0.9 are identified as proton candidates;
charged kaon candidates are required to have $\mathcal{R}(K|\pi)$ $>$ 0.9
and $\mathcal{R}(K|p)$ $>$ 0.4; and charged pion candidates to
have $\mathcal{R}(\pi|K)$ $>$ 0.4 and
$\mathcal{R}(\pi|p)$ $>$ 0.4. A likelihood ratio for electron
identification, $\mathcal{R}(e)$, is formed from ACC, CDC, and ECL information~\cite{eid},
and is required to be less than 0.9 for all charged tracks to suppress electrons. For each charged track, the impact
parameters with respect to the IP are required to be less than 0.1~cm and 2.0~cm perpendicular to,
and along the $e^+$ beam direction, respectively. The $\Lambda_c^+$ candidate is reconstructed
by combining $p$, $K^{-}$, and $\pi^+$ candidates. A vertex fit is performed for
the reconstructed $\Lambda_c^+$ candidate with a requirement of $\chi_{\rm vertex}^2 <40$.
The required $x_p$ value of $\Lambda_c^+ \to pK^-\pi^+$ is the same as that of
corresponding signal channel.

For the $\Xi_c^0 \to \Xi^- \pi^+$ reconstruction, tracks having
$\mathcal{R}(\pi|K)$ $>$ 0.6 and
$\mathcal{R}(\pi|p)$ $>$ 0.6 are identified as pion candidates.
For the $\pi^+$ that is the direct daughter of the $\Xi_c^0$, the impact parameters with respect to the IP
are required to be less than 0.5~cm and 4.0~cm perpendicular to, and along the $e^+$ beam direction,
respectively, and the transverse momentum is restricted to be higher than 0.1~GeV/$c$. The $\Lambda$ candidates
are selected using the same procedure as in the signal channel. The $\Xi^-$ candidates are
reconstructed from the combinations of selected $\Lambda$ and $\pim$ candidates. We define the $\Xi^-$ signal region as
$|M(\Lambda\pim)-m(\Xi^-)|<$ 6.5~MeV/$c^2$ ($\sim$3.0$\sigma$). Finally, the reconstructed $\Xi^-$ candidate is
combined with a $\pip$ to form the $\Xi_c^0$ candidate. We perform vertex fits for the $\Lambda$,
$\Xi^-$, and $\Xi_c^0$ candidates. To suppress the combinatorial backgrounds, we require the flight directions
of $\Lambda$ and $\Xi^-$ candidates, which are reconstructed from their fitted production and decay vertices,
to be within $5^\circ$ of their momentum directions. The requirement on $x_p$ value of $\Xi_c^0 \to \Xi^- \pip$
is the same as that of the corresponding signal channel.

\section{\boldmath Extractions of $\Lambda_c^+ \to \Sigma^+ \gamma$ and $\Xi_c^0 \to \Xi^0 \gamma$ signal yields}
\begin{figure*}[htbp]
	\begin{center}
		\includegraphics[width=8.8cm]{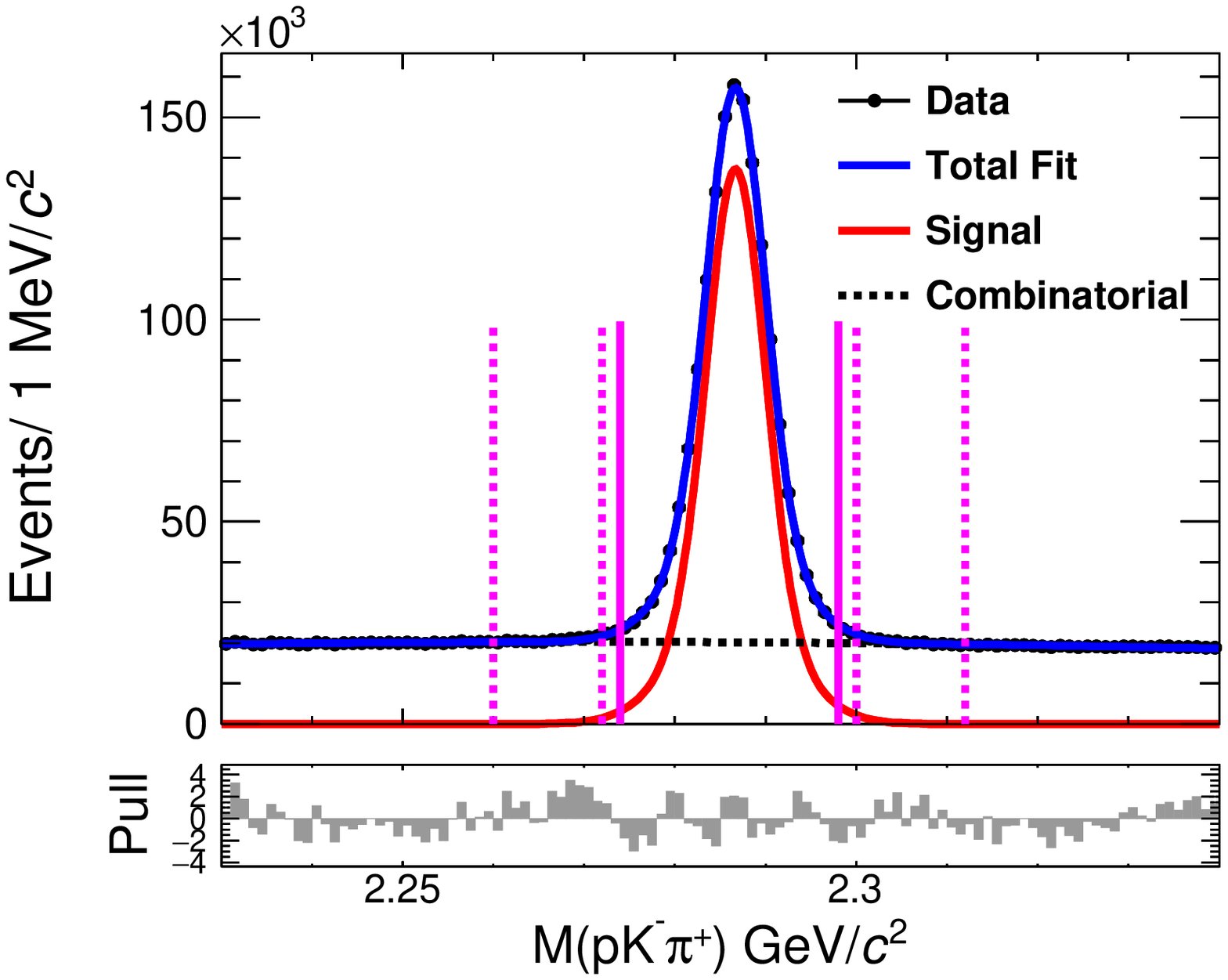}
		\includegraphics[width=8.8cm]{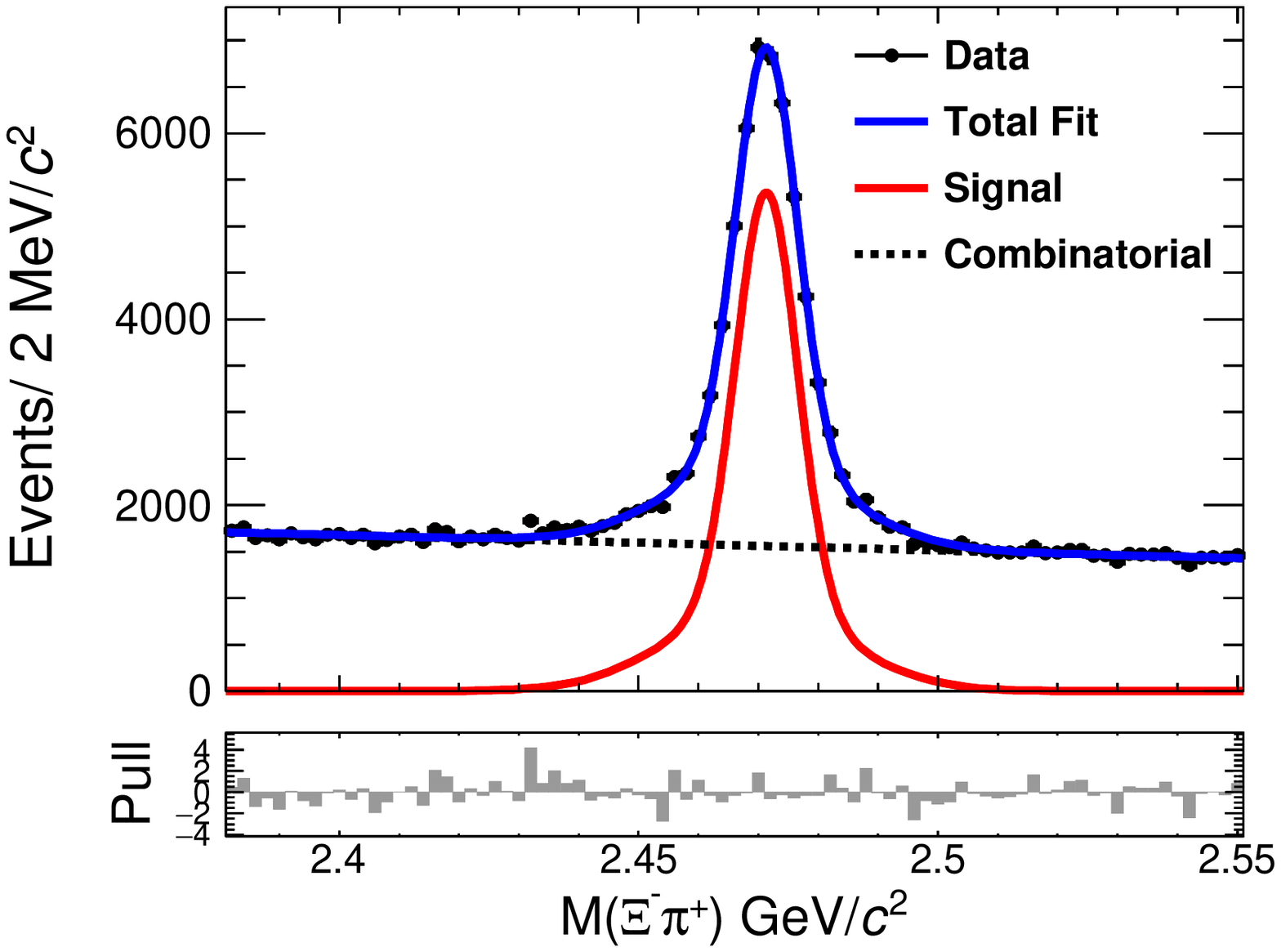}
		\put(-440,145){\bf (a)} \put(-185,145){\bf (b)}
		\caption{The invariant mass distributions of (a) $p K^- \pip$ and (b) $\Xi^- \pip$ from the reconstructed
			$\Lambda_c^+ \to p K^- \pi^+ $ and $\Xi_c^0 \to \Xi^- \pi^+$ candidates in data. The points with
			error bars represent the data, the blue solid curves show the best-fit results, the red solid curves
			denote the fitted signals, and the black dashed curves represent the fitted combinatorial backgrounds.
		    In (a), the pink solid lines indicate the required signal region, and the pink dashed lines
		    denote the defined sideband regions.}\label{Fig2}
	\end{center}
\end{figure*}

The $M(pK^-\pip)$ and $M(\Xi^- \pip)$ distributions of events corresponding to the normalization channels
$\Lambda_c^+ \to p K^- \pi^+$ and $\Xi_c^0 \to \Xi^- \pi^+$  in data are shown in Figs.~\ref{Fig2}(a) and~\ref{Fig2}(b),
respectively. To extract the $\Lambda_c^+ \to p K^- \pi^+ $ and $\Xi_c^0 \to \Xi^- \pi^+$ signal yields,
we perform binned extended maximum-likelihood fits to the $M(pK^-\pip)$ and $M(\Xi^-\pi^+)$ distributions.
In the fits, double-Gaussian functions are taken as the signal probability density functions (PDFs) for the
$\Lambda_c^+$ and $\Xi_c^0$, and the combinatorial background PDFs are parametrized by a second-order polynomial for the $M(pK^-\pip)$ distribution and a
first-order polynomial for $M(\Xi^-\pi^+)$. The parameters of the signal and combinatorial background PDFs are free.
The fitted results are displayed in Fig.~\ref{Fig2} along with the distributions of the pulls $(N_{\rm data}-N_{\rm fit})/\sigma_{\rm data}$,
where $\sigma_{\rm data}$ is the uncertainty on $N_{\rm data}$, and the fitted signal yields
of $\Lambda_c^+ \to p K^- \pip$ and $\Xi_c^0 \to \Xi^- \pip$ decays in data are ($1281910 \pm 2040$)
and ($45063 \pm 445$), respectively.

\begin{figure}[htbp]
	\begin{center}
		\includegraphics[width=9cm]{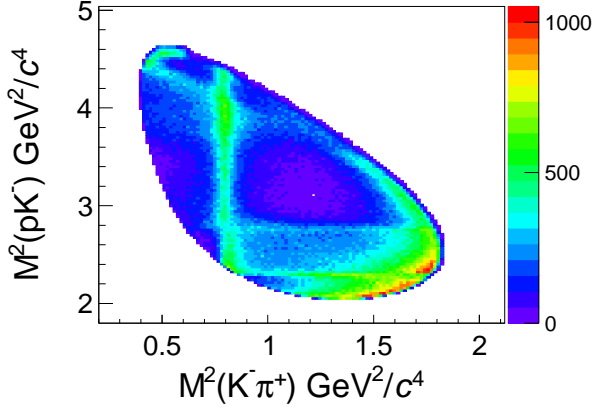}\hspace{1cm}
		\caption{Dalitz distribution of the reconstructed $\Lambda_c^+ \to p K^- \pi^+$ candidates in data.}\label{Fig3}
	\end{center}
\end{figure}

For the three-body decay $\Lambda_c^+ \to p K^- \pip$, the reconstruction efficiency can vary across
the phase space, as visualized in a Dalitz distribution~\cite{Dalitz:1953cp}.
Figure~\ref{Fig3} shows the Dalitz distribution of $M^2(p K^-)$ versus $M^2(K^-\pip)$ from data
in the $\Lambda_c^+$ signal region with the normalized $\Lambda_c^+$ mass sidebands subtracted.
The signal region of $\Lambda_c^+$ is defined as 2.274~GeV/$c^2$ $<$ $M(p K^{-} \pip)$ $<$ 2.298~GeV/$c^2$,
and the sideband regions are defined as 2.260~GeV/$c^2$ $<$ $M(p K^{-} \pip)$ $<$ 2.272~GeV/$c^2$
or 2.300~GeV/$c^2$ $<$ $M(p K^{-} \pip)$ $<$ 2.312~GeV/$c^2$. We divide the Dalitz distribution into $120\times120$ bins,
with a bin size of 0.027~$\rm GeV^2$/$c^4$ for $M^2(pK^{-})$ and 0.016~$\rm GeV^2$/$c^4$ for $M^2(K^-\pip)$.
The reconstruction efficiency averaged over the Dalitz distribution is calculated by the
formula $\epsilon = \Sigma_{i}s_i/\Sigma_{j}(s_j/\epsilon_{j})$~\cite{Belle:2021vyq}, where $i$ and $j$ run over all bins;
$s_{i/j}$ and $\epsilon_{j}$ are the number of signal events in data and the reconstruction efficiency from signal MC simulation for each bin, respectively.
The reconstruction efficiency for each bin is obtained by dividing the number of signal events after applying
the selection criteria with the normalized sidebands subtracted by the number of generated events.
The corrected reconstruction efficiency for $\Lambda_c^+ \to p K^- \pip$
is determined to be $(12.79 \pm 0.02)$\%. For the two-body decay $\Xi_c^0 \to \Xi^- \pip$, we estimate the reconstruction
efficiency directly from the simulated events by the ratio $n_{\rm sel}$/$n_{\rm gen}$, where $n_{\rm sel}$
and $n_{\rm gen}$ are the numbers of true signal events surviving the selection criteria and generated events,
respectively. The signal reconstruction efficiency for $\Xi_c^0 \to \Xi^- \pip$ is determined to be $(16.96 \pm 0.05)$\%.

\begin{figure*}[htbp]
	\begin{center}
		\includegraphics[width=8.8cm]{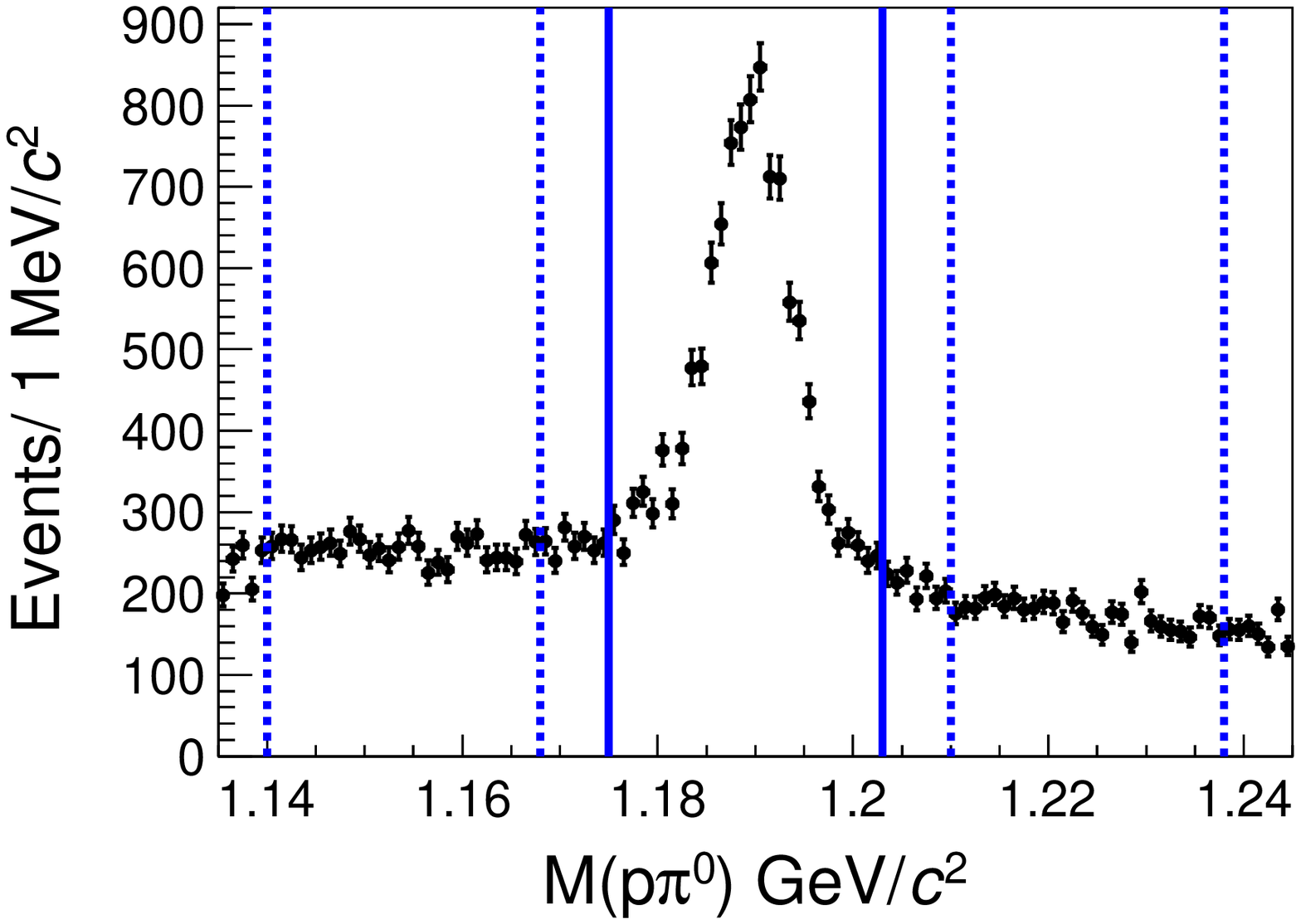}
		\includegraphics[width=8.8cm]{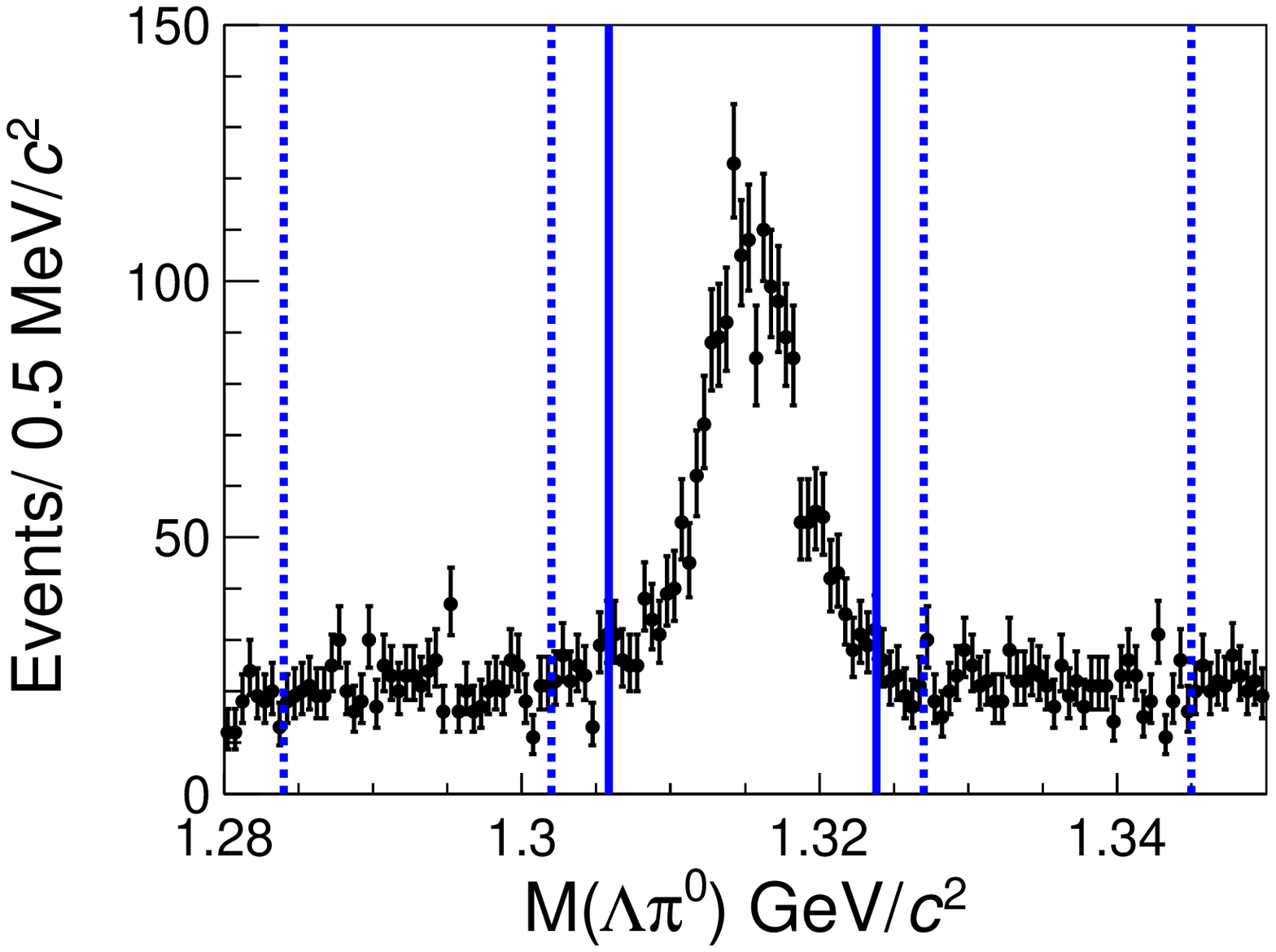}
		\put(-440,145){\bf (a)} \put(-185,145){\bf (b)}
		\caption{The invariant mass distributions of (a) $p \pi^0$ and (b) $\Lambda \pi^0$ from the reconstructed
			$\Lambda_c^+ \to \Sigma^+ \gamma$ and $\Xi_c^0 \to \Xi^0 \gamma$ candidates in the $\Lambda_c^+$ and
			$\Xi_c^0$ signal regions in data, respectively. The points with error bars represent the data,
			the blue solid lines indicate the required signal regions, and the blue dashed lines
			denote the defined sidebands.}\label{Fig4}
	\end{center}
\end{figure*}

After applying the event selection criteria mentioned in Sec.~\ref{secII}, the invariant mass
distributions of $p \pi^0$ and $\Lambda\pi^0$ from the reconstructed
$\Lambda_c^+ \to \Sigma^+ \gamma$ and $\Xi_c^0 \to \Xi^0 \gamma$ candidates
in data are shown in Figs.~\ref{Fig4}(a) and \ref{Fig4}(b), respectively.
There are significant $\Sigma^+$ and $\Xi^0$ signals observed in the $\Lambda_c^+$ and $\Xi_c^0$
signal regions, respectively. The signal regions of $\Sigma^+$ and $\Xi^0$ candidates are defined as
$|M(p\pi^0) - m(\Sigma^+)|$ $<$ 14~MeV/$c^2$ ($\sim$2.5$\sigma$) and $|M(\Lambda \pi^0) - m(\Xi^0)|$ $<$
9~MeV/$c^2$ ($\sim$2.5$\sigma$). We define the $\Sigma^+$ and $\Xi^0$ sideband regions as
1.140~GeV/$c^2$ $<$ $M(p\pi^0)$ $<$ 1.168~GeV/$c^2$ or 1.210~GeV/$c^2$ $<$ $M(p\pi^0)$ $<$ 1.238~GeV/$c^2$,
and 1.284~GeV/$c^2$ $<$ $M(\Lambda \pi^0)$ $<$ 1.302~GeV/$c^2$ or 1.327~GeV/$c^2$ $<$ $M(\Lambda \pi^0)$ $<$
1.345~GeV/$c^2$, respectively, which are twice as wide as the corresponding signal regions.
The blue solid lines indicate the required $\Sigma^+$ and $\Xi^0$ signal regions, and the blue dashed lines
represent the defined $\Sigma^+$ and $\Xi^0$ sideband regions.

\begin{figure*}[htbp]
	\begin{center}
		\includegraphics[width=8.8cm]{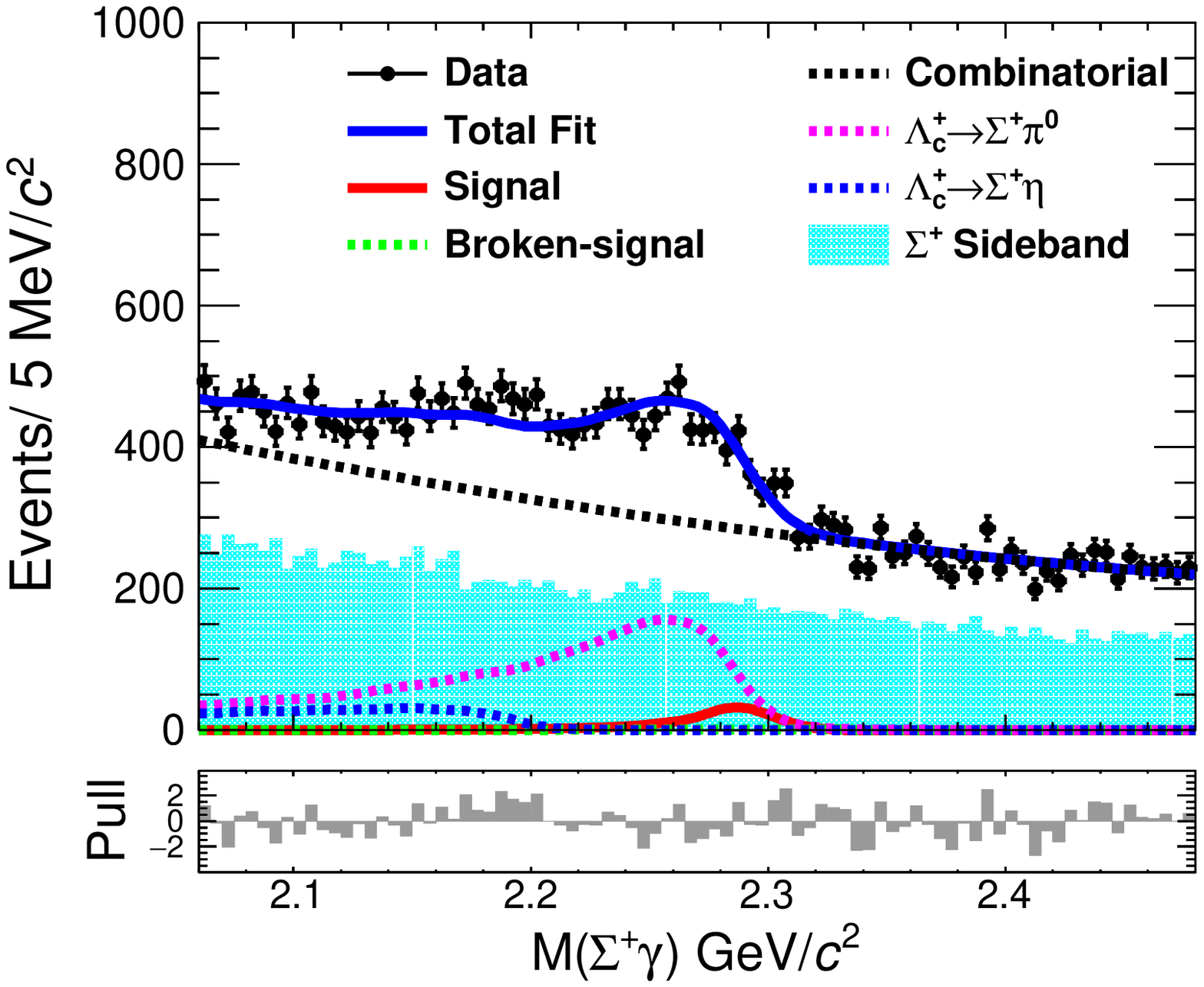}
		\includegraphics[width=8.8cm]{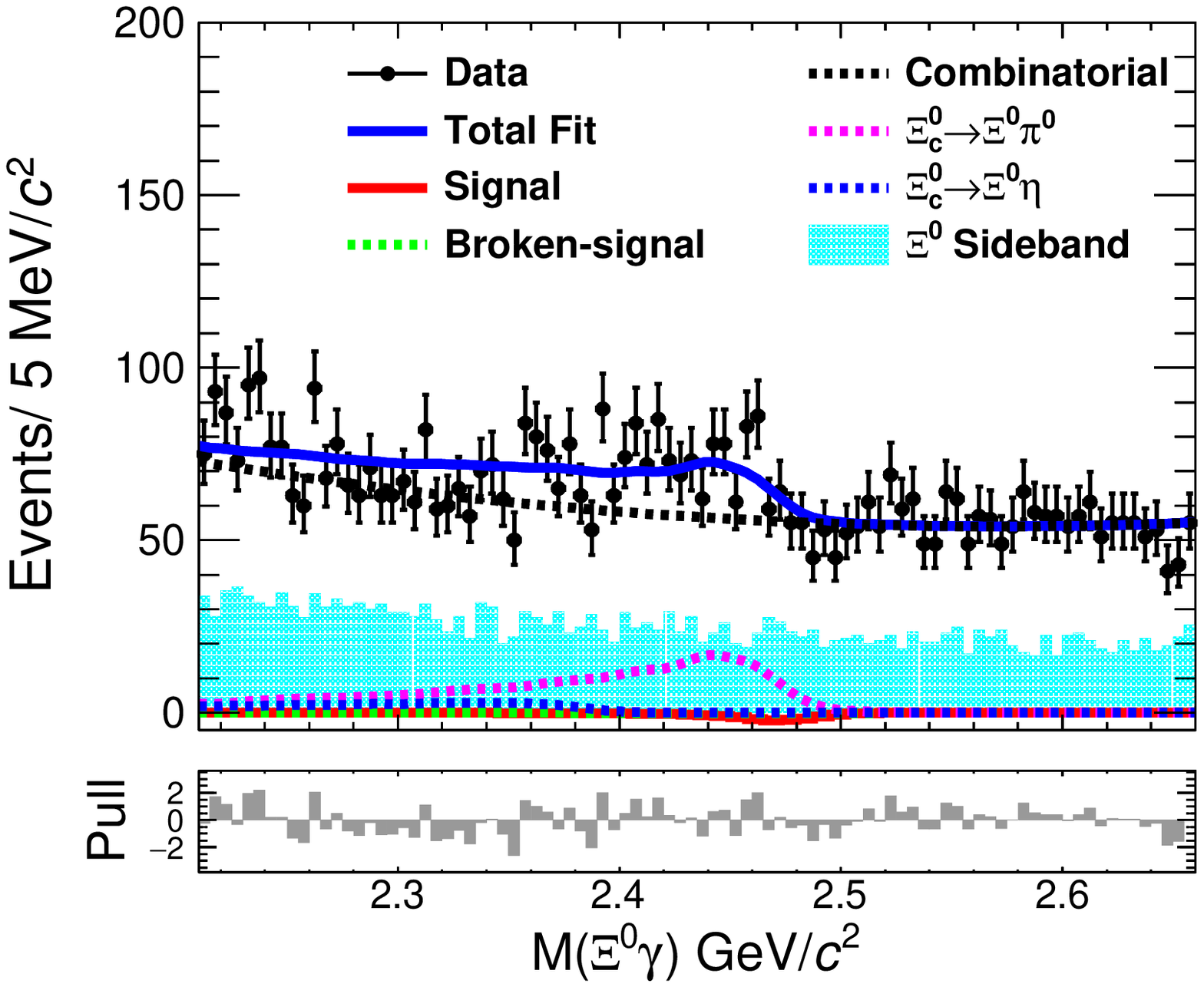}
		\put(-445,153){\bf (a)} \put(-193,153){\bf (b)}
		\caption{The invariant mass distributions of (a) $\Sigma^+ \gamma$ and (b) $\Xi^0 \gamma$ from the reconstructed $\Lambda_c^+ \to \Sigma^+ \gamma$ and
			$\Xi_c^0 \to \Xi^0 \gamma$ candidates in data. The points with error bars represent the data, and the cyan shaded histograms denote events from the normalized $\Sigma^+$ and $\Xi^0$ sidebands. The blue solid curves show the best-fit results. The red solid and green dashed curves indicate the fitted signal and broken-signal components. The black dashed curves are the fitted combinatorial backgrounds. The pink and blue dashed curves show the fitted
			peaking backgrounds from the contributions of $\Lambda_c^+ \to \Sigma^+\pi^0(\to \gamma \gamma)/\Xi_c^0 \to \Xi^0\pi^0(\to \gamma \gamma)$
			and $\Lambda_c^+ \to \Sigma^+\eta(\to \gamma \gamma)/\Xi_c^0 \to \Xi^0\eta(\to \gamma \gamma)$, respectively.}\label{Fig5}
	\end{center}
\end{figure*}

Figures~\ref{Fig5}(a) and \ref{Fig5}(b) display the invariant mass spectra of $\Sigma^+ \gamma$ and
$\Xi^0 \gamma$ from data, and the cyan shaded histograms represent events from the normalized
$\Sigma^+$ and $\Xi^0$ sidebands, respectively. There are broad peaking backgrounds found in
both $M(\Sigma^+ \gamma)$ and $M(\Xi^0 \gamma)$ distributions in data and inclusive MC samples.
According to a study of inclusive MC samples using the TopoAna package~\cite{topo},
we found that these peaking backgrounds in the $M(\Sigma^+ \gamma)$ and $M(\Xi^0 \gamma)$
distributions arise from the contributions of $\Lambda_c^+ \to \Sigma^+ \pi^0(\to \gamma\gamma)$ and $\Sigma^+ \eta (\to \gamma\gamma)$
, and $\Xi_c^0 \to \Xi^0 \pi^0(\to \gamma\gamma)$ and $\Xi^0 \eta(\to \gamma\gamma)$ decays respectively,
where one of the two photons has been missed.

For the $\Lambda_c^+ \to \Sigma^+ \gamma$ mode, the expected
peaking background events from the contributions of $\Lambda_c^+ \to \Sigma^+ \pi^0(\to \gamma \gamma)$ and
$\Lambda_c^+ \to \Sigma^+ \eta(\to \gamma \gamma)$ in the $M(\Sigma^+ \gamma)$ distribution
are estimated according to the formulae
\begin{displaymath}
	\begin{aligned}
	N_{\Sigma^+ \pi^0}^{\Sigma^+ \gamma} = &~ \eff_{\Sigma^+ \pi^0}^{\Sigma^+ \gamma} \times \frac{N_{p K^- \pip}^{\rm obs}\BR(\Lambda_c^+ \to \Sigma^+ \pi^0)}{\eff_{p K^{-} \pip}} \\ &~\times \frac{\BR(\Sigma^+ \to p \pi^0)\BR(\pi^0 \to \gamma \gamma)\BR(\pi^0 \to \gamma\gamma)}{\BR(\Lambda_c^+ \to p K^- \pi^+)}
   \end{aligned}
\end{displaymath}
and
\begin{displaymath}
	\begin{aligned}
	N_{\Sigma^+ \eta}^{\Sigma^{+} \gamma} = &~ \eff_{\Sigma^+ \eta}^{\Sigma^+ \gamma} \times \frac{N_{p K^- \pip}^{\rm obs}\BR(\Lambda_c^+ \to \Sigma^+ \eta)}{\eff_{p  K^{-} \pip}}  \\ &~ \times \frac{\BR(\Sigma^+ \to p \pi^0)\BR(\eta \to \gamma \gamma)\BR(\pi^0 \to \gamma\gamma)}{\BR(\Lambda_c^+ \to p K^- \pi^+)},
   \end{aligned}
\end{displaymath}
where $\eff_{\Sigma^+ \pi^0}^{\Sigma^+ \gamma}$ = 0.36\% and $\eff_{\Sigma^+ \eta}^{\Sigma^+ \gamma}$ = 0.46\%
are the reconstruction efficiencies of $\Lambda_c^+ \to \Sigma^+ \pi^0(\to \gamma \gamma)$ and
$\Lambda_c^+ \to \Sigma^+ \eta(\to \gamma\gamma)$ decays under the $\Lambda_c^+ \to \Sigma^+ \gamma$ selection criteria
obtained by signal MC simulations; $\epsilon_{p K^- \pi^+}$ = $(12.79 \pm 0.02)$\% denotes the reconstruction efficiency of
$\Lambda_c^+ \to p K^- \pi^+$ decay; $N^{\rm obs}_{p K^- \pip}$ = $(1281910 \pm 2040)$
represents the observed $\Lambda_c^+ \to p K^- \pi^+$ signal events in data; the branching fractions
$\BR(\Lambda_c^+ \to \Sigma^+ \pi^0)$, $\BR(\Lambda_c^+ \to \Sigma^+ \eta)$,
$\BR(\Lambda_c^+ \to p K^- \pi^+)$, $\BR(\Sigma^+ \to p \pi^0)$, $\BR(\pi^0 \to \gamma \gamma)$, and $\BR(\eta \to \gamma\gamma)$
are taken from the Particle Data Group (PDG)~\cite{PDG}. Using the values above, the peaking background events from
the contributions of $\Lambda_c^+ \to \Sigma^+ \pi^0(\to \gamma \gamma)$ and $\Lambda_c^+ \to \Sigma^+ \eta(\to \gamma \gamma)$
in the $M(\Sigma^+ \gamma)$ distribution are determined to be $N_{\Sigma^+ \pi^0}^{\Sigma^+ \gamma} = (3617 \pm 344)$ and
$N_{\Sigma^{+} \eta}^{\Sigma^{+} \gamma} = (649 \pm 297)$, respectively, where the uncertainties
are mainly from the input branching fractions.

For the $\Xi_c^0 \to \Xi^0 \gamma$ mode, the corresponding backgrounds can not
be estimated using this method, because the branching fractions of
$\Xi_c^0 \to \Xi^0 \pi^0$ and $\Xi_c^0 \to \Xi^0 \eta$ decays have not yet been measured. To estimate the numbers of
peaking background events from the contributions of $\Xi_c^0 \to \Xi^0 \pi^0(\to \gamma \gamma)$ and
$\Xi_c^0 \to \Xi^0 \eta(\to \gamma \gamma)$
in the $M(\Xi^0 \gamma)$ distribution, we study the background channels $\Xi_c^0 \to \Xi^0 \pi^0$
and $\Xi^0 \eta$ in data. The $\Xi^0$ candidates are selected using the same criteria used for the signal
mode $\Xi_c^0 \to \Xi^0 \gamma$. The $\pi^0$ $(\eta)$ candidates
are reconstructed from two photons having energy exceeding 110~MeV (230~MeV) in both barrel and endcaps of the
ECL, and are required to have momentum exceeding 800~MeV/$c$ (900~MeV/$c$) in the laboratory frame.
Mass-constrained fits are performed for $\pi^0$ and $\eta$ candidates with a requirement of $\chi^2 <5$. For
$\eta \to \gamma \gamma$ reconstruction, $\ppp(\pi^0) < 0.6$ is required. The $\Xi^0 \pi^0$ and $\Xi^0 \eta$ combinations are then
made to form $\Xi_c^0$ candidates, and the scaled momentum $x_p > 0.55$ is required.

\begin{figure*}[htbp]
	\begin{center}
		\includegraphics[width=8.8cm]{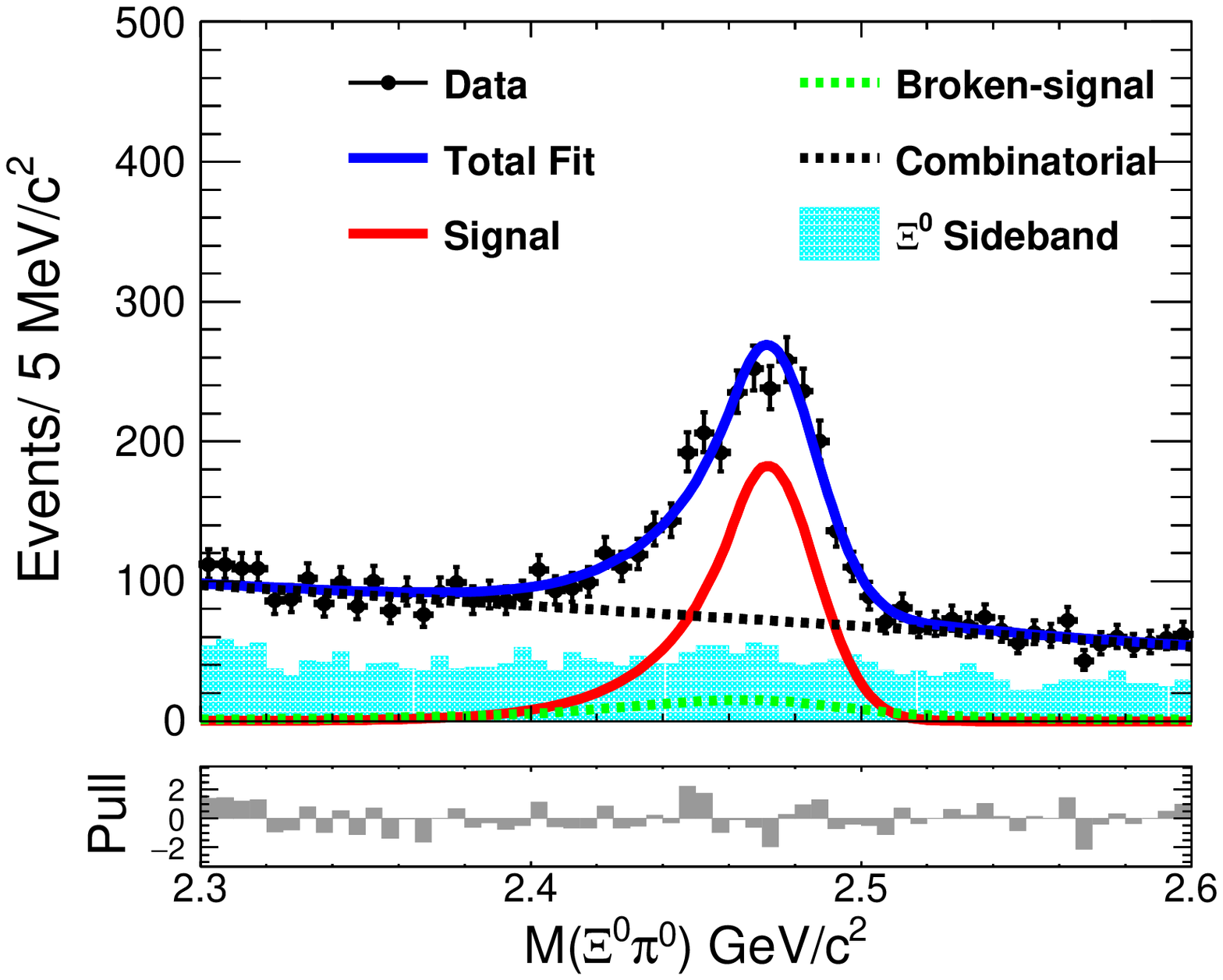}
		\includegraphics[width=8.8cm]{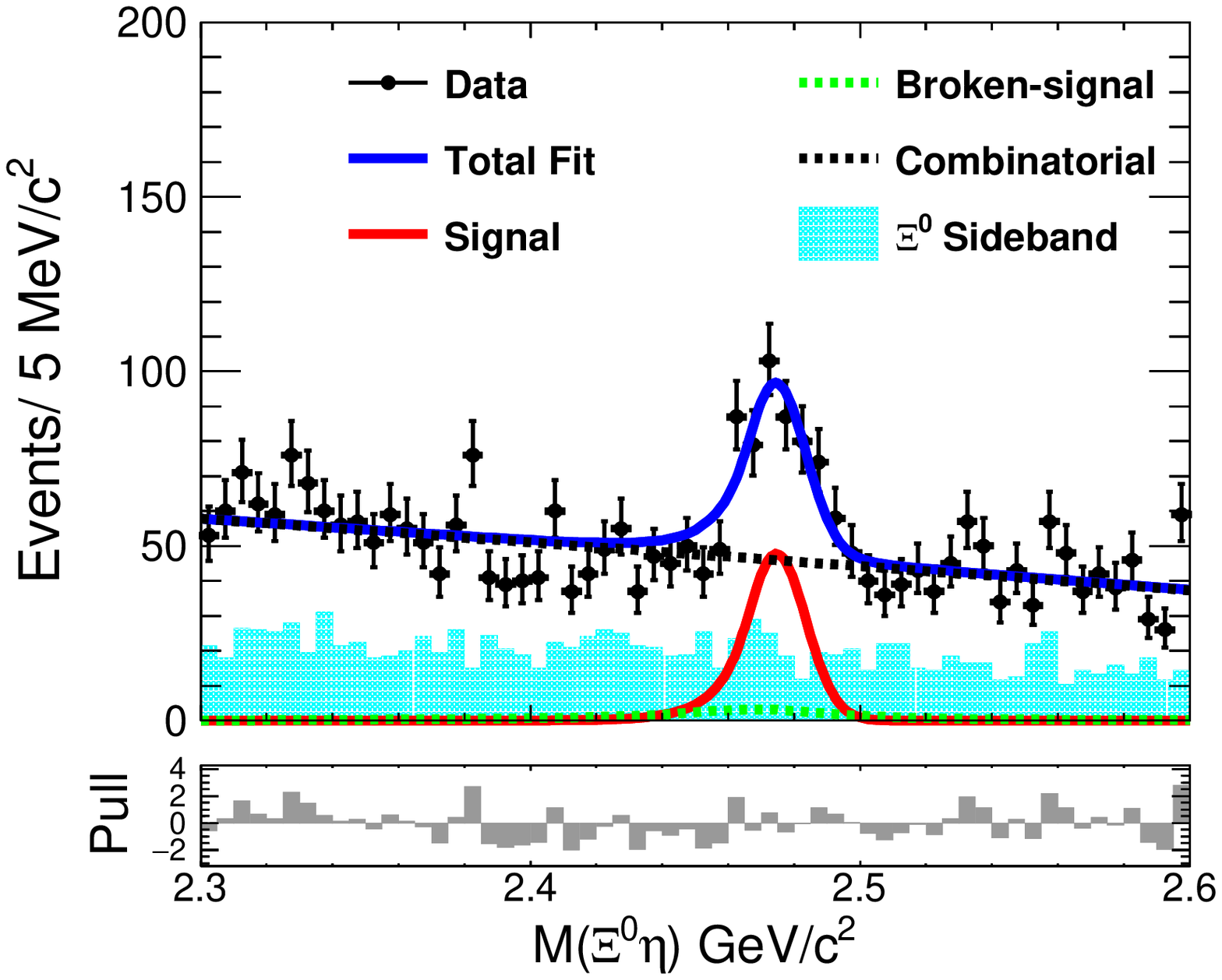}
		\put(-445,153){\bf (a)} \put(-193,153){\bf (b)}
		\caption{The invariant mass distributions of (a) $\Xi^0 \pi^0$ and (b) $\Xi^0 \eta$ from reconstructed $\Xi_c^0 \to \Xi^0 \pi^0(\to \gamma\gamma)$ and $\Xi_c^0 \to \Xi^0 \eta(\to \gamma\gamma)$  candidates in data. The points with error bars represent the data, and the cyan shaded histograms denote events from the normalized $\Xi^0$ sidebands. The blue solid curves show the best-fit results. The red solid and green dashed curves indicate the fitted signal and broken-signal components. The black dashed curves are the fitted combinatorial backgrounds.}\label{Fig6}
	\end{center}
\end{figure*}

After applying the event selection criteria above, the invariant mass spectra of $\Xi^0 \pi^0$ and $\Xi^0 \eta$ in data are shown in
Figs.~\ref{Fig6}(a) and \ref{Fig6}(b), respectively. The cyan shaded histograms represent events from the normalized $\Xi^0$ sidebands,
where the defined sideband regions of $\Xi^0$  are the same as those of the signal channel.
We observe significant $\Xi_c^0 \to \Xi^0 \pi^0(\to \gamma\gamma)$
and $\Xi_c^0 \to \Xi^0 \eta(\to \gamma\gamma)$ signals in data, and no evident peaking backgrounds are found in the normalized $\Xi^0$ sidebands
or in the inclusive MC samples.
To extract the $\Xi_c^0$ signal yields from the $\Xi_c^0 \to \Xi^0 \pi^0(\to \gamma\gamma)$ and $\Xi_c^0 \to \Xi^0 \eta(\to\gamma\gamma)$
decays, we perform unbinned extended maximum-likelihood fits to the $M(\Xi^0 \pi^0)$ and $M(\Xi^0\eta)$ distributions. The likelihood
function includes the following components: (1) The signal PDF of $\Xi_c^0$ candidates. (2) The broken-signal PDF from
true $\Xi_c^0 \to \Xi^0 \pi^0(\to \gamma\gamma)$ or  $\Xi_c^0 \to \Xi^0 \eta(\to \gamma\gamma)$ signal decays,
where at least one of the final state particle candidates is wrongly assigned in reconstruction. (3) The combinatorial background PDF.
A Crystal-Ball (CB) function~\cite{CB} is taken as the $\Xi_c^0$ signal PDF, the broken-signal is represented by a non-parametric
(multi-dimensional) kernel estimation PDF~\cite{Cranmer:2000du} based on the signal MC simulation, and the combinatorial background PDF is described by a
first-order polynomial. The parameters of signal and background PDFs are free.
The ratios of signal to broken-signal components are fixed to (83.0\%:17.0\%) and (81.2\%:18.8\%) for
$\Xi_c^0 \to \Xi^0 \pi^0(\to \gamma\gamma)$ and $\Xi_c^0 \to \Xi^0 \eta(\to \gamma\gamma)$, respectively, according to the signal MC simulations. Hereinafter, the broken-signal component is regarded as part of the signal. Figure~\ref{Fig6} displays
the fit results along with the pull distributions. The fitted $\Xi_c^0 \to \Xi^0 \pi^0(\to \gamma\gamma)$
and $\Xi_c^0 \to \Xi^0 \eta(\to \gamma\gamma)$ signal yields in data are ($1940 \pm 78$) and ($288 \pm 33$), respectively.

The peaking background events from the contributions of $\Xi_c^0 \to \Xi^0 \pi^0(\to \gamma \gamma)$ and $\Xi_c^0 \to \Xi^0 \eta(\to \gamma \gamma)$
in the $M(\Xi^0 \gamma)$ distribution are estimated from the following formulae
\begin{displaymath}
	\begin{aligned}
    N_{\Xi^0 \pi^0}^{\Xi^0\gamma} = \frac{N_{\Xi^0\pi^0}}{\epsilon_{\Xi^0\pi^0}} \times \epsilon_{\Xi^0\pi^0}^{\Xi^0\gamma} \notag
   \end{aligned}
\end{displaymath}
and
\begin{displaymath}
	\begin{aligned}
 	N_{\Xi^0 \eta}^{\Xi^0\gamma} = \frac{N_{\Xi^0\eta}}{\epsilon_{\Xi^0\eta}} \times \epsilon_{\Xi^0\eta}^{\Xi^0\gamma}. \notag
	\end{aligned}
\end{displaymath}
Here, $N_{\Xi^0 \pi^0} = (1940 \pm 78)$ and $N_{\Xi^0 \eta} = (288 \pm 33)$ are the observed $\Xi_c^0 \to \Xi^0 \pi^0(\to \gamma \gamma)$
and $\Xi_c^0 \to \Xi^0 \eta(\to \gamma \gamma)$ signal events under $\Xi_c^0 \to \Xi^0 \pi^0(\to \gamma \gamma)$ and
$\Xi_c^0 \to \Xi^0 \eta(\to \gamma \gamma)$ selection criteria in data with reconstruction efficiencies of
$\epsilon_{\Xi^0\pi^0} = (2.14 \pm 0.01 )\%$ and $\epsilon_{\Xi^0\eta} = (2.61 \pm 0.01)\%$, respectively;
$\epsilon_{\Xi^0\pi^0}^{\Xi^0\gamma} = (0.35 \pm 0.01)\%$ and $\epsilon_{\Xi^0\eta}^{\Xi^0\gamma} = (0.56 \pm 0.01)\%$
are the reconstruction efficiencies of $\Xi_c^0 \to \Xi^0 \pi^0(\to \gamma \gamma)$ and $\Xi_c^0 \to \Xi^0 \eta(\to \gamma \gamma)$
under $\Xi_c^0 \to \Xi^0 \gamma$ selection criteria, respectively.
Using the values above, the peaking background events from the contributions of
$\Xi_c^0 \to \Xi^0 \pi^0(\to \gamma \gamma)$ and $\Xi_c^0 \to \Xi^0 \eta(\to \gamma \gamma)$
in the $M(\Xi^0 \gamma)$ distribution are  determined to be $N_{\Xi^0 \pi^0}^{\Xi^0 \gamma} = (317 \pm 13)$ and
$N_{\Xi^0 \eta}^{\Xi^0 \gamma} = (62 \pm 7)$, respectively, where the uncertainties are statistical only.

To extract the signal yields of $\Lambda_c^+ \to \Sigma^+ \gamma$ and $\Xi_c^0 \to \Xi^0 \gamma$ decays, we perform
unbinned extended maximum-likelihood fits to $M(\Sigma^+ \gamma)$ and $M(\Xi^0 \gamma)$ distributions in Fig.~\ref{Fig5}, and the following
components are included in the fits: (1) The signal PDF ($\F_{\rm S}$) of $\Lambda_c^+$ or $\Xi_c^0$ candidates.
(2) The broken-signal PDF ($\F_{\rm BS}$) from the true $\Lambda_c^+ \to \Sigma^+ \gamma$ or $\Xi_c^0 \to \Xi^0 \gamma$
signal decay, where at least one of the final state particle candidates is wrongly assigned in reconstruction.
(3) The peaking background PDFs ($\F_{\pi^0}$ and $\F_{\eta}$) from the contributions of
$\Lambda_c^+ \to \Sigma^+\pi^0(\to \gamma\gamma)/\Xi_c^0 \to \Xi^0 \pi^0(\to \gamma\gamma)$
and $\Lambda_c^+ \to \Sigma^+ \eta (\to \gamma\gamma)/\Xi_c^0 \to \Xi^0 \eta(\to \gamma\gamma)$.
(4) Other combinatorial background PDF ($\F_{\rm B}$). The likelihood function is defined as
\begin{displaymath}
	\begin{aligned}
	\mathcal{L} =  & \frac{e^{-(n_{\rm S} + n_{\rm B_{1}} + n_{\rm B_{2}})}}{N!}\prod_{i}^{N}\{n_{\rm S}[f_{\rm S} \F_{\rm S}(M_i) + (1-f_{\rm S}) \F_{\rm BS}(M_i)] \\
	& + n_{\rm B_{1}}[f_{\rm B_{1}}\F_{\pi^0}(M_i) + (1-f_{\rm B_1})\F_{\eta}(M_i)] + n_{\rm B_{2}}\F_{\rm B}(M_i)\} \notag,
  \end{aligned}
\end{displaymath}
where $N$ is the total number of observed events; $n_{\rm S}$, $n_{\rm B_1}$, and $n_{\rm B_2}$ are
the total numbers of signal events (the broken-signal is regarded as part of the signal), the peaking background events,
and other combinatorial background events, respectively; $f_{\rm S}$ and $f_{\rm B_1}$ indicate
the fractions of the correctly reconstructed signal events in the total number of signal events and the peaking
background events from $\Lambda_c^+ \to \Sigma^+ \pi^0(\to\gamma\gamma)$ or $\Xi_c^0 \to \Xi^0\pi^0(\to\gamma\gamma)$
in the total number of peaking  background events, respectively; $M$ represents the $\Sigma^+ \gamma$ or $\Xi^0 \gamma$
invariant mass; $i$ denotes the event index. The signal PDF is parameterized by a CB function~\cite{CB}, and the
parameters of the signal PDF are fixed to those obtained from signal MC simulation. The broken-signal
and the peaking background are represented by
non-parametric (multi-dimensional) kernel estimation PDFs~\cite{Cranmer:2000du} based on signal MC simulations. The other
combinatorial background PDF is a second-order polynomial. The values of
$n_{\rm S}$, $n_{\rm B_{1}}$, and $n_{\rm B_{2}}$ are free in the fit. The ratios of signal
to broken-signal components are fixed to (91.3\%:8.7\%) and (84.7\%:15.3\%) for $\Lambda_c^+ \to \Sigma^+ \gamma$
and $\Xi_c^0 \to \Xi^0 \gamma$, respectively, according to the signal MC simulations.
Since the shapes of the two peaking background components cannot be separated well,
the ratios of peaking background events from $\Lambda_c^+ \to \Sigma^+ \pi^0(\to \gamma\gamma)$ relative
to those from $\Lambda_c^+ \to \Sigma^+ \eta(\to \gamma\gamma)$
and peaking background events from $\Xi_c^0 \to \Xi^0 \pi^0(\to \gamma\gamma)$ relative to those from $\Xi_c^0 \to \Xi^0 \eta(\to \gamma\gamma)$
are fixed to (84.8\%:15.2\%) and (83.6\%:16.4\%), respectively, based on the expected events in the $M(\Sigma^+\gamma)$
and $M(\Xi^0 \gamma)$ distributions in data. The fitted results are displayed in Fig.~\ref{Fig5}
along with the pull distributions, and the fitted signal yields of $\Lambda_c^+ \to \Sigma^+ \gamma$ and
$\Xi_c^0 \to \Xi^0 \gamma$ decays in data are ($340 \pm 110$) and ($-18 \pm48 $), respectively. The fitted
numbers of peaking background events are consistent with the corresponding expected numbers of events within the
ranges of uncertainties. The statistical significance of the $\Lambda_c^+ \to \Sigma^+ \gamma$ decay is 3.2$\sigma$ calculated using
$\sqrt{-2\ln(\mathcal{L}_{0}/\mathcal{L}_\text{max})}$, where $\mathcal{L}_{0}$ and $\mathcal{L}_{\rm max}$ are the
likelihoods of the fits without and with signal and broken-signal components, respectively.
To estimate the signal significance of $\Lambda_c^+ \to \Sigma^+ \gamma$ decay after considering
the systematic uncertainties, several alternative fits discussed in the section of systematic uncertainty to the $M(\Sigma^+ \gamma)$ spectrum are performed.
The signal significance of $\Lambda_c^+ \to \Sigma^+ \gamma$ decay is larger than 2.2$\sigma$ in all cases.	
We take 2.2$\sigma$ as the signal significance with systematic uncertainties included.

\section{\boldmath Systematic Uncertainties \label{secV}}
The systematic uncertainties on the measurements of branching fraction ratios
can be divided into two categories as discussed below.

The sources of multiplicative systematic uncertainties include detection-efficiency-related
uncertainties, the modeling of MC event generation, and branching fractions of intermediate states.
Note that the uncertainties from detection-efficiency-related sources partially cancel in
the ratio to the normalization channel.

The detection-efficiency-related uncertainties include those from tracking efficiency,
PID efficiency, $\pi^0$ reconstruction efficiency, and photon reconstruction efficiency.
Based on a study of $D^{*+} \to \pi^+ D^0(\to K_S^0 \pi^+ \pi^-)$
decay, the tracking efficiency uncertainty is evaluated to be 0.35\% per track.
Using $D^{*+} \to D^0 \pip$, $D^0 \to K^{-}\pip$,
and $\Lambda \to p \pim$ control samples, the PID efficiency uncertainties are estimated to be 0.95\%
per kaon and 0.96\% per pion for the normalization channel $\Lambda_c^+ \to p K^- \pip$.
For the normalization channel $\Xi_c^0 \to \Xi^-(\to\Lambda\pim)\pip$,
the PID efficiency uncertainties of $\pi^+$ from the $\Xi_c^0$ decay and $\pi^-$ from the $\Xi^-$ decay
are considered separately, because the $\pi^+$ has a higher momentum. The PID efficiency ratio between the data
and MC simulation of $\pi^+$ is found to be $\epsilon_{\rm data}/\epsilon_{\rm MC}$ = ($95.4 \pm 0.7$)\%,
so we take 95.4\% and 0.7\% as the efficiency correction factor and PID uncertainty for $\pip$; the
PID efficiency ratio between the data and MC simulation of $\pi^-$ is found to be
$\epsilon_{\rm data}/\epsilon_{\rm MC}$ = ($99.5 \pm 0.8$)\%, and 1.3\% is taken as the PID uncertainty of $\pim$.
The uncertainties from proton PID efficiency and $\Lambda$
reconstruction mostly cancel in the ratio with the normalization channel.
The PID uncertainties of $K$ and $\pi$ are added linearly to obtain the final PID uncertainties,
which are 1.9\% and 2.0\% for the measurements of
$\BR(\Lambda_c^+ \to \Sigma^+ \gamma)/\BR(\Lambda_c^+ \to p K^- \pip)$ and
$\BR(\Xi_c^0 \to \Xi^0\gamma)/\BR(\Xi_c^0 \to \Xi^- \pip)$, respectively.
The uncertainties associated with $\pi^0$ and radiative photon reconstruction
efficiencies are treated as independent, and are estimated to be 2.3\%~\cite{pi0Un} and
2.0\%~\cite{gamUn}, respectively.
Assuming these uncertainties are independent and adding them in quadrature,
the final detection-efficiency-related uncertainties are obtained, as listed in Table~\ref{tab:errsum}.

We assume that both $\Lambda_c^+ \to \Sigma^+ \gamma$ and $\Xi_c^0 \to \Xi^0 \gamma$ decays
are isotropic in the rest frame of the parent particle, and a phase space model is used to
generate signal events by default. Alternative angular distributions $(1\pm\cos^2\theta_\gamma)$
are also generated, where $\theta_\gamma$ is the angle between the $\gamma$ momentum vector
and the boost direction from the laboratory frame in the $\Lambda_c^+$ or $\Xi_c^0$ C.M.\ frame.
The maximum differences in the reconstruction efficiencies between the alternatives and default signal MC samples
are taken as systematic uncertainties, which are 6.9\% and 2.3\% for
$\Lambda_c^+ \to \Sigma^+ \gamma$ and $\Xi_c^0 \to \Xi^0 \gamma$, respectively.

For the measurement of $\BR(\Lambda_c^+ \to \Sigma^+ \gamma)/\BR(\Lambda_c^+ \to p K^- \pip)$, the
uncertainties from $\BR(\Sigma^+ \to p \pi^0)$ and $\BR(\pi^0 \to \gamma \gamma)$ are 0.6\% and 0.035\%~\cite{PDG},
which are added in quadrature as the total uncertainty from branching fractions of intermediate states.
For the measurement of $\BR(\Xi_c^0 \to \Xi^0 \gamma)/\BR(\Xi_c^0 \to \Xi^- \pip)$,
the uncertainties from $\BR(\Xi^- \to \Lambda \pi^-)$, $\BR(\Xi^0 \to \Lambda \pi^0)$,
and $\BR(\pi^0 \to \gamma\gamma)$  are 0.035\%, 0.012\%, and 0.035\%~\cite{PDG},
which are added in quadrature as the total uncertainty from branching fractions of intermediate states.

Additive systematic uncertainties associated with the combinatorial background PDF, fit range,
$\Lambda_c^+$ or $\Xi_c^0$ mass resolution, the ratio of the signal component to the broken-signal
component, and the peaking backgrounds from the contributions of
$\Lambda_c^+ \to \Sigma^+\pi^0(\to \gamma \gamma)/\Xi_c^0 \to \Xi^0\pi^0(\to \gamma \gamma)$
and $\Lambda_c^+ \to \Sigma^+\eta(\to \gamma \gamma)/\Xi_c^0 \to \Xi^0\eta(\to \gamma \gamma)$
are considered as follows: (1) The combinatorial background PDF is replaced by a higher- or
lower-order polynomial. (2) The fit range is changed by $\pm$30~MeV/$c^2$. (3) To consider the uncertainty
associated with $\Lambda_c^+$ or $\Xi_c^0$ mass resolution, the signal PDF of $\Lambda_c^+$ or $\Xi_c^0$
is replaced by a Gaussian function with free resolution convolved with the fixed signal shape from signal MC simulation.
(4) Since the ratio of the signal component to the broken-signal component in signal MC
simulation may not be consistent with that in data, we add an extra broken-signal component described by the same PDF as original broken-signal
with free yield to fit the data. (5) The number of peaking background events from the contributions of $\Lambda_c^+ \to \Sigma^+\pi^0/\Xi_c^0 \to \Xi^0\pi^0$
and $\Lambda_c^+ \to \Sigma^+\eta/\Xi_c^0 \to \Xi^0\eta$ is constrained with a Gaussian function whose mean
value and width are equal to the number of expected events in $M(\Sigma^+ \gamma)$ or $M(\Xi^0 \gamma)$ distribution
and the corresponding uncertainty. For the normalization channels $\Lambda_c^+ \to p K^{-} \pip$ and $\Xi_c^0 \to \Xi^- \pip$,
the additive systematic uncertainties associated with the background PDF and fit range are estimated using the same method
as above, and then summed in quadrature to obtain the total additive systematic uncertainties, which are
1.9\% and 3.1\% for $\Lambda_c^+ \to p K^{-} \pip$ and $\Xi_c^0 \to \Xi^- \pip$, respectively.

Since no evident $\Lambda_c^+ \to \Sigma^+ \gamma$ or $\Xi_c^0 \to \Xi^0 \gamma$ signals are
found, the upper limits on the numbers of signal events ($N^{\rm UL}$) at 90\% credibility level (C.L.)\
are determined by solving the equation
\begin{displaymath}
	\begin{aligned}
		\int_0^{N^{\rm UL}} \mathcal{L} (N) dN /\int_0^{+\infty}\mathcal{L} (N) dN = 0.9, \notag
	\end{aligned}
\end{displaymath}
where $N$ represents the assumed $\Lambda_c^+ \to \Sigma^+ \gamma$ or $\Xi_c^0 \to \Xi^0 \gamma$
signal events and the $\mathcal{L} (N)$ is the corresponding maximized likelihood of the fit to the assumption,
and the systematic uncertainties are taken into account in two steps. First, when we study
the additive systematic uncertainties described above, we calculate the upper limit for each possible case,
and take the most conservative upper limit at 90\% C.L. on the number of signal events. For the
$\Lambda_c^+ \to \Sigma^+ \gamma$ decay, when the combinatorial background PDF is replaced by a third-order
polynomial, the fit range is reduced by 30~MeV/$c^2$, and the signal PDF of $\Lambda_c^+$  is replaced by a
Gaussian function with free resolution convolved with the fixed signal shape, the obtained upper limit is
the most conservative. For the $\Xi_c^0 \to \Xi^0 \gamma$ decay, when the fit range is reduced by 30~MeV/$c^2$,
the signal PDF of $\Xi_c^0$  is replaced by a Gaussian function with free resolution convolved with the fixed signal shape,
and the number of peaking background events is constrained with a Gaussian function, the obtained upper
limit is the most conservative. Then, the multiplicative systematic uncertainties from the signal and normalization
channels and the additive systematic uncertainty from the normalization channel are summed in quadrature to give the total systematic uncertainty,
and the likelihood with the most conservative upper limit is convolved with a Gaussian
function whose width is equal to the corresponding total systematic uncertainty. Furthermore, to obtain the upper
limits on the absolute branching fractions $\BR(\Lambda_c^+ \to \Sigma^+ \gamma)$ and $\BR(\Xi_c^0 \to \Xi^0 \gamma)$
at 90\% C.L., the multiplicative systematic uncertainties from the signal and normalization channels, the additive
systematic uncertainty from the normalization channel, and the uncertainty from branching fraction
$\BR(\Lambda_c^+ \to p K^- \pip)$ or $\BR(\Xi_c^0 \to \Xi^- \pip)$ are added in quadrature as total systematic
uncertainty, and then the likelihood with the most conservative upper limit is convolved with a Gaussian function
whose width equals to the corresponding total systematic uncertainty.

Assuming all the sources are independent and adding the multiplicative systematic uncertainties and
additive systematic uncertainties from normalization channel in quadrature, the total systematic uncertainties
are obtained. All the systematic uncertainties are summarized in Table~\ref{tab:errsum}.

\begin{table}[htbp]
	\caption{\label{tab:errsum} Relative systematic uncertainties (\%) on the measurements of branching fraction ratios $\BR(\Lambda_c^+ \to \Sigma^+ \gamma)/\BR(\Lambda_c^+ \to p K^- \pip)$ and $\BR(\Xi_c^0 \to \Xi^0 \gamma)/\BR(\Xi_c^0 \to \Xi^-\pip)$.}
	\begin{tabular}{lcc}
		\hline\hline
		\multicolumn{1}{c}{Sources} & \multicolumn{1}{c}{$\frac{\BR(\Lambda_c^+ \to \Sigma^+ \gamma)}{\BR(\Lambda_c^+ \to p K^{-} \pip)}$}  & \multicolumn{1}{c}{$\frac{\BR(\Xi_c^0 \to \Xi^0 \gamma)}{\BR(\Xi_c^0 \to \Xi^{-}\pip)}$} \\
		\hline
		Detection efficiency             & 3.7  &  3.7   \\
		Branching fraction               & 0.6  &  0.1   \\	
		The modeling of MC events        & 6.9  &  2.3   \\
		Additive uncertainty             & 1.9  &  3.1   \\
		Sum                              & 8.1  &  5.4   \\
		\hline\hline
	\end{tabular}
\end{table}

\section{\boldmath Calculations of the ratios of branching fractions}
The most conservative upper limits on the numbers of $\Lambda_c^+ \to \Sigma^+ \gamma$
and $\Xi_c^0 \to \Xi^0 \gamma$ signal events at 90\% C.L.\ are determined
to be 608 and 91, respectively, and then the upper limits at 90\% C.L.\ on the
ratios of the branching fractions are determined from the following formulae
\begin{displaymath}
	\begin{aligned}
		\frac{\BR(\Lambda_c^+ \to \Sigma^+ \gamma)}{\BR(\Lambda_c^+ \to p K^- \pip)}
		< &~ \frac{N^{\rm UL}_{\Sigma^+ \gamma} \epsilon_{p K^- \pip}}{N^{\rm obs}_{p K^- \pip} \epsilon_{\Sigma^+ \gamma}} \\ &
		\times \frac{1}{\BR(\Sigma^+ \to p \pi^0) \BR(\pi^0 \to \gamma \gamma)} \\
		= &~4.0 \times 10^{-3} \notag
	\end{aligned}
\end{displaymath}
and
\begin{displaymath}
	\begin{aligned}
		\frac{\BR(\Xi_c^0 \to \Xi^0 \gamma)}{\BR(\Xi_c^0 \to \Xi^- \pip)}
		< &~ \frac{N^{\rm UL}_{\Xi^0 \gamma}\epsilon_{\Xi^- \pip}}
		{N^{\rm obs}_{\Xi^- \pip}\epsilon_{\Xi^0 \gamma}} \\ &
		\times 	\frac{\BR(\Xi^- \to \Lambda \pi^-)}	{\BR(\Xi^0 \to \Lambda \pi^0) \BR(\pi^0 \to \gamma \gamma)}\\
		= &~1.2 \times 10^{-2} \notag.
	\end{aligned}
\end{displaymath}	
Here $N^{\rm UL}_{\Sigma^+ \gamma}$ and $N^{\rm UL}_{\Xi^0 \gamma}$ represent the upper limits on the numbers of
$\Lambda_c^+ \to \Sigma^+ \gamma$ and $\Xi_c^0 \to \Xi^0 \gamma$ signal events at 90\% C.L.; $N^{\rm obs}_{p K^{-} \pip}$
and $N^{\rm obs}_{\Xi^-\pip}$ denote the observed signal events of $\Lambda_c^+ \to p K^- \pip$ and $\Xi_c^0 \to \Xi^- \pip$ decays
in data; $\epsilon_{\Sigma^+ \gamma}$, $\epsilon_{\Xi^0 \gamma}$, $\epsilon_{p K^{-} \pip}$, and $\epsilon_{\Xi^- \pip}$ are the corresponding reconstruction efficiencies, which are obtained from the signal MC simulations and are listed in Table~\ref{tab:summary};
the branching fractions $\BR(\Sigma^+ \to p \pi^0)$ = ($51.57 \pm 0.03$)\%, $\BR(\pi^0 \to \gamma \gamma)$ = ($98.823 \pm 0.034$)\%,
$\BR(\Xi^- \to  \Lambda \pim)$ = ($99.887 \pm 0.035$)\%, and $\BR(\Xi^0 \to \Lambda \pi^0)$ = ($99.524 \pm 0.012$)\%
are taken from the PDG~\cite{PDG}. Since the statistical significance of $\Lambda_c^+ \to \Sigma^+ \gamma$
is 3.2$\sigma$, we also give the ratio of branching fractions with $N^{\rm obs}_{\Sigma^+ \gamma} = (340 \pm 110)$
replacing $N^{\rm UL}_{\Sigma^+ \gamma}$ in above formula,
$\BR(\Lambda_c^+ \to \Sigma^+ \gamma)/\BR(\Lambda_c^+ \to pK^-\pip) = (2.23 \pm 0.72 \pm 0.63) \times 10^{-3}$,
where the first uncertainty is statistical, and the second one arises from the multiplicative
and additive systematic uncertainties discussed in Sec.~\ref{secV}.

\begin{table}[htbp]
	\caption{\label{tab:summary} Summary of the fitted signal events ($N^{\rm obs}$), the upper limits at 90\% C.L.\
		on the numbers of signal events ($N^{\rm UL}$), and the reconstruction efficiencies ($\epsilon$). All the uncertainties
		here are statistical only.}
	\begin{tabular}{lr@{$\pm$}lcr@{$\pm$}l}
		\hline\hline
		\multicolumn{1}{c}{Modes} & \multicolumn{2}{c}{$N^{\rm obs}$} & \multicolumn{1}{c}{$N^{\rm UL}$} & \multicolumn{2}{c}{$\epsilon$(\%)}\\
		\hline
		$\Lambda_c^+ \to \Sigma^+ \gamma$  &          340&110     & 608  &  2.98&0.01 \\
		$\Xi_c^0 \to \Xi^0 \gamma$         &        $-$18&48      &  91  &  3.03&0.01 \\	
		$\Lambda_c^+ \to p K^- \pip$       &      1281910&2040    & ...  & 12.79&0.02 \\	
		$\Xi_c^0 \to \Xi^-   \pip$         &        45063&445     & ...  & 16.96&0.05 \\	
		\hline\hline
	\end{tabular}
\end{table}

\section{\boldmath Summary}
In summary, using the entire data sample of 980~fb$^{-1}$ integrated luminosity collected by the Belle detector,
we perform the first search for the weak radiative decays $\Lambda_c^+ \to \Sigma^+ \gamma$ and $\Xi_c^0 \to \Xi^0 \gamma$.
No evidences for $\Lambda_c^+ \to \Sigma^+ \gamma$ or $\Xi_c^0 \to \Xi^0 \gamma$ signals are found. The upper limits at 90\% C.L.\ on the
ratios of the branching fractions
\begin{displaymath}
	\begin{aligned}
   \frac{\BR(\Lambda_c^+ \to \Sigma^+ \gamma)}{\BR(\Lambda_c^+ \to p K^- \pi^+)} < 4.0 \times 10^{-3} \notag
   \end{aligned}
\end{displaymath}
and
\begin{displaymath}
	\begin{aligned}
    \frac{\BR(\Xi_c^0 \to \Xi^0 \gamma)}{\BR(\Xi_c^0 \to \Xi^- \pi^+)} < 1.2 \times 10^{-2} \notag
   \end{aligned}
\end{displaymath} are measured. Taking $\BR(\Lambda_c^+ \to p K^- \pi^+) = (6.28 \pm 0.32)\%$ and
$\BR(\Xi_c^0 \to \Xi^- \pi^+) = (1.43 \pm 0.32)\%$, we determine the upper limits at 90\% C.L.\ on
the absolute branching fractions $\BR(\Lambda_c^+ \to \Sigma^+ \gamma) < 2.6 \times 10^{-4}$ and
$\BR(\Xi_c^0 \to \Xi^0 \gamma) < 1.8 \times 10^{-4}$.
The measured upper limits on the absolute branching fractions of $\Lambda_c^+ \to \Sigma^+ \gamma$ and
$\Xi_c^0 \to \Xi^0 \gamma$ are slightly smaller than the theoretical predictions of case (II) in Ref.~\cite{Uppal},
which naively considered the flavor dependence of charmed baryon wave-function squared at the origin $|\psi(0)|^{2}$.

\section{\boldmath ACKNOWLEDGMENTS}

This work, based on data collected using the Belle detector, which was
operated until June 2010, was supported by 
the Ministry of Education, Culture, Sports, Science, and
Technology (MEXT) of Japan, the Japan Society for the 
Promotion of Science (JSPS), and the Tau-Lepton Physics 
Research Center of Nagoya University; 
the Australian Research Council including grants
DP180102629, 
DP170102389, 
DP170102204, 
DE220100462, 
DP150103061, 
FT130100303; 
Austrian Federal Ministry of Education, Science and Research (FWF) and
FWF Austrian Science Fund No.~P~31361-N36;
the National Natural Science Foundation of China under Contracts
No.~11675166,  
No.~11705209;  
No.~11975076;  
No.~12135005;  
No.~12175041;  
No.~12161141008; 
Key Research Program of Frontier Sciences, Chinese Academy of Sciences (CAS), Grant No.~QYZDJ-SSW-SLH011; 
Project ZR2022JQ02 supported by Shandong Provincial Natural Science Foundation;
the Ministry of Education, Youth and Sports of the Czech
Republic under Contract No.~LTT17020;
the Czech Science Foundation Grant No. 22-18469S;
Horizon 2020 ERC Advanced Grant No.~884719 and ERC Starting Grant No.~947006 ``InterLeptons'' (European Union);
the Carl Zeiss Foundation, the Deutsche Forschungsgemeinschaft, the
Excellence Cluster Universe, and the VolkswagenStiftung;
the Department of Atomic Energy (Project Identification No. RTI 4002) and the Department of Science and Technology of India; 
the Istituto Nazionale di Fisica Nucleare of Italy; 
National Research Foundation (NRF) of Korea Grant
Nos.~2016R1\-D1A1B\-02012900, 2018R1\-A2B\-3003643,
2018R1\-A6A1A\-06024970, RS\-2022\-00197659,
2019R1\-I1A3A\-01058933, 2021R1\-A6A1A\-03043957,
2021R1\-F1A\-1060423, 2021R1\-F1A\-1064008, 2022R1\-A2C\-1003993;
Radiation Science Research Institute, Foreign Large-size Research Facility Application Supporting project, the Global Science Experimental Data Hub Center of the Korea Institute of Science and Technology Information and KREONET/GLORIAD;
the Polish Ministry of Science and Higher Education and 
the National Science Center;
the Ministry of Science and Higher Education of the Russian Federation, Agreement 14.W03.31.0026, 
and the HSE University Basic Research Program, Moscow; 
University of Tabuk research grants
S-1440-0321, S-0256-1438, and S-0280-1439 (Saudi Arabia);
the Slovenian Research Agency Grant Nos. J1-9124 and P1-0135;
Ikerbasque, Basque Foundation for Science, Spain;
the Swiss National Science Foundation; 
the Ministry of Education and the Ministry of Science and Technology of Taiwan;
and the United States Department of Energy and the National Science Foundation.
These acknowledgements are not to be interpreted as an endorsement of any
statement made by any of our institutes, funding agencies, governments, or
their representatives.
We thank the KEKB group for the excellent operation of the
accelerator; the KEK cryogenics group for the efficient
operation of the solenoid; and the KEK computer group and the Pacific Northwest National
Laboratory (PNNL) Environmental Molecular Sciences Laboratory (EMSL)
computing group for strong computing support; and the National
Institute of Informatics, and Science Information NETwork 6 (SINET6) for
valuable network support.

\end{document}